\PassOptionsToPackage{table,xcdraw}{xcolor}
\documentclass[11pt,a4paper]{article}
\pdfoutput=1
\usepackage{jheppub}
\usepackage{gensymb}
\DeclareSymbolFont{matha}{OML}{txmi}{m}{it}
\DeclareMathSymbol{\varv}{\mathord}{matha}{118}
\usepackage{amsmath}
\usepackage{amssymb}
\usepackage{graphicx}
\usepackage{subfigure}
\usepackage{pstricks}
\usepackage{bm}
\usepackage{ulem}
\usepackage{pbox}
\usepackage{placeins}
\usepackage{booktabs}
\usepackage[T1]{fontenc}
\usepackage{footnote}
\usepackage{pdfpages}
\usepackage{hhline}
\usepackage{multirow}
\usepackage{multicol}
\usepackage{enumitem}
\usepackage[toc,page]{appendix}
\allowdisplaybreaks
\usepackage{comment}
\usepackage{float}
\usepackage{slashed}
\usepackage{xcolor}
\usepackage{textcomp}
\usepackage{gensymb}
\usepackage{multirow}
\usepackage[numbers]{natbib}
\usepackage{notoccite}
\usepackage{rotating}
\usepackage{tabularx}
\usepackage{multirow}
\usepackage{xurl}
\usepackage[labelfont=bf]{caption} 

\FloatBarrier
\usepackage[many]{tcolorbox}

\renewcommand{\thetable}{\Roman{table}}

\newcommand{\eq}[1]{Eq.~\eqref{#1}}

\definecolor{MyDarkBlue}{rgb}{0.1, 0.1, 0.8} 
\definecolor{MyLightBlue}{rgb}{0.22,0.51,0.9}
\definecolor{MyGreen}{rgb}{0.0, 0.5, 0.0}
\definecolor{BrickRed}{rgb}{0.8, 0.25, 0.33}

\RequirePackage{hyperref}
\hypersetup{colorlinks, citecolor=MyGreen,linkcolor=cyan, urlcolor=MyLightBlue}

\makeatletter
\gdef\@fpheader{}
\makeatother
\begin{document}

\title{A flavor-inspired radiative neutrino mass model
}

\author[a]{\bf J. Julio,}
\author[b]{\bf Shaikh Saad,}
\author[c]{and \bf Anil Thapa}

\affiliation[a]{National Research and Innovation Agency, Kompleks Puspiptek Serpong, South Tangerang 15314, Indonesia}

\affiliation[b]{Department of Physics, University of Basel, Klingelbergstrasse\ 82, CH-4056 Basel, Switzerland}

\affiliation[c]{Department of Physics, University of Virginia, Charlottesville, Virginia 22904-4714, USA}

\emailAdd{julio@brin.go.id, shaikh.saad@unibas.ch, wtd8kz@virginia.edu}
\abstract{
One of the most important discoveries in particle physics is the observation of nonzero neutrino masses, which dictates that the Standard Model (SM) is incomplete. Moreover, several pieces of evidence of lepton flavor universality violation (LFUV), gathered in the last few years, hint toward physics beyond the SM. TeV-scale scalar leptoquarks  are the leading candidates for explaining these flavor anomalies in semileptonic charged and neutral current B-decays, the muon, and the electron magnetic dipole moments that can also participate in neutrino mass generation. In this work, we hypothesize that neutrino masses and LFUV have a common new physics origin and propose a new two--loop neutrino mass model that has the potential to resolve some of these flavor anomalies via leptoquarks and offers rich phenomenology. After deriving the neutrino mass formula for this newly-proposed model, we perform a detailed numerical analysis focusing on neutrino and charged lepton flavor violation phenomenology, where the latter provides stringent constraints on the Yukawa couplings and leptoquark masses. Finally, present and future bounds on the model's parameter space are scrutinized with exemplified benchmark scenarios. 
}

\maketitle
\section{Introduction}
The observation of neutrino oscillations was the first direct hint that the Standard Model of particle physics is imperfect and must be extended. Lepton flavor universality (LFU), a solid prediction of the SM,  can be easily violated in the beyond SM (BSM) models, where the particles preferentially couple to certain generations of leptons.  In the last several years, indications of LFU violation (LFUV) have been observed in both $b\to s\ell\ell$ and $b\to c\ell\nu$ processes. Observables associated with these transitions are the well-known $R_{K^{(\ast)}}$ and $R_{D^{(\ast)}}$ ratios, respectively. LFU in the SM predicts the former ratio to be unity with uncertainties less than $1\%$. A deficit in this neutral-current transition has been observed consistently over the years in several experiments, and LHCb recently updated their measurements \cite{LHCb:2021trn} that increased the significance of the deviation. Moreover, with yet  another observed deviation in $Br\left(B^0_s\to \mu^+\mu^-  \right)$ \cite{LHCb:2015wdu,LHCb:2017rmj,ATLAS:2018cur,CMS:2019bbr},  the combined significance of the deviation is uplifted to $4.7\sigma$. On the other hand, the $R_{D^{(\ast)}}$ ratio  differs from unity due to the  substantial mass difference between tauon and muon. An enhancement of this charged-current
transition is reported by several experimental measurements, which combinedly leads to approximately $3\sigma$ deviation from the SM value \cite{Na:2015kha, Aoki:2016frl}.

Besides, there has been a longstanding tension between the theoretical prediction of the anomalous magnetic dipole moment (AMDM) of the muon $(g-2)_\mu$ and the value measured at the BNL E821 experiment  \cite{Bennett:2006fi}. The FNAL E989 experiment \cite{Abi:2021gix} has recently announced its result, which has a smaller uncertainty and is fully compatible with  the previous best measurement. Together, these two experiments  show a remarkably large deviation with a significance of $4.2\sigma$ with respect to the theory prediction \cite{Aoyama:2020ynm}. Various new physics models are proposed to explain the observed significant  departure. For a most recent review see Ref.~\cite{Athron:2021iuf}.  The SM prediction given in Ref.~\cite{Aoyama:2020ynm} is based on the estimate of the leading-order hadronic vacuum
polarization contribution, evaluated from a data-driven approach. On the other hand, if recent lattice computations~\cite{Borsanyi:2020mff,Ce:2022kxy,Alexandrou:2022amy} are considered, then the tension reduces to $1.5\sigma$ from $4.2\sigma$. However, if these new lattice results hold, they point towards a large $\sim 4.2\sigma$ discrepancy with the low-energy $e^+e^-\to$ hadrons cross-section data with respect to SM predictions~\cite{Keshavarzi:2020bfy,Crivellin:2020zul,DiLuzio:2021uty,Ce:2022eix}.

On top of that, the electron AMDM $(g-2)_e$ is also measured in the experiments with an unprecedented level of accuracy. Recently, improved measurement  \cite{Parker:2018vye}  of the fine-structure constant utilizing Caesium atom shows a $-2.4\sigma$ deviation in comparison with the direct experimental measurement~\cite{Hanneke:2008tm}.  Lately, these anomalies in the lepton AMDMs have gained a lot of attention in the theory community; for simultaneous explains of the muon and the electron AMDMs in various BSM frameworks, see e.g. Refs.~\cite{Giudice:2012ms, Davoudiasl:2018fbb,Crivellin:2018qmi,Liu:2018xkx,Dutta:2018fge, Han:2018znu,Crivellin:2019mvj,Endo:2019bcj, Abdullah:2019ofw, Bauer:2019gfk,Badziak:2019gaf,Hiller:2019mou,CarcamoHernandez:2019ydc,Cornella:2019uxs,Endo:2020mev,CarcamoHernandez:2020pxw,Haba:2020gkr, Bigaran:2020jil, Jana:2020pxx,Calibbi:2020emz,Chen:2020jvl,Yang:2020bmh,Hati:2020fzp,Dutta:2020scq,Botella:2020xzf,Chen:2020tfr, Dorsner:2020aaz, Arbelaez:2020rbq, Jana:2020joi,Chua:2020dya,Chun:2020uzw,Li:2020dbg,DelleRose:2020oaa,Kowalska:2020zve,Hernandez:2021tii,Bodas:2021fsy,Cao:2021lmj,Mondal:2021vou,CarcamoHernandez:2021iat,Han:2021gfu,Escribano:2021css,CarcamoHernandez:2021qhf,Chang:2021axw,Chowdhury:2021tnm,Bharadwaj:2021tgp,Borah:2021khc,Bigaran:2021kmn,Jana:2021jjm,Li:2021wzv,Biswas:2021dan,Barman:2021xeq,Chowdhury:2022jde}.  It is noteworthy to point out that a more recent measurement of fine-structure constant utilizing Rubidium atom~\cite{Morel:2020dww} shows somewhat consistent with the direct measurement of $a_e$~\cite{Hanneke:2008tm}. This new result~\cite{Morel:2020dww} finds $\Delta a_e= +1.6\sigma$, indicating a $\sim 5\sigma$  disagreement between these two experiments (\cite{Parker:2018vye} and \cite{Morel:2020dww}). Therefore,  the electron $g-2$ situation requires clarification from future experiments.

All these flavor anomalies mentioned above are strongly pointing toward physics beyond the SM. Interestingly, the prime candidates to solve these flavor anomalies are leptoquarks (LQs), i.e., hypothetical particles that combine the properties of leptons and quarks (for a recent review on LQs, see Ref.~\cite{Dorsner:2016wpm}). The existence of LQs are highly motivated since particles of this type are naturally predicted by Grand Unified Theories. Explanation of flavor anomalies~\cite{Dorsner:2013tla, Sakaki:2013bfa, Duraisamy:2014sna,  Hiller:2014yaa, Buras:2014fpa,Gripaios:2014tna, Freytsis:2015qca, Pas:2015hca, Bauer:2015knc,Fajfer:2015ycq,  Deppisch:2016qqd, Li:2016vvp, Becirevic:2016yqi,Becirevic:2016oho, Sahoo:2016pet, Bhattacharya:2016mcc, Duraisamy:2016gsd, Barbieri:2016las,Crivellin:2017zlb, DAmico:2017mtc,Hiller:2017bzc, Becirevic:2017jtw, Cai:2017wry,Alok:2017sui, Sumensari:2017mud,Buttazzo:2017ixm,Crivellin:2017dsk, Guo:2017gxp,Aloni:2017ixa,Assad:2017iib, DiLuzio:2017vat,Calibbi:2017qbu,Chauhan:2017uil,Cline:2017aed,Sumensari:2017ovu, Biswas:2018jun,Muller:2018nwq,Blanke:2018sro, Schmaltz:2018nls,Azatov:2018knx, Sheng:2018vvm, Becirevic:2018afm, Hati:2018fzc, Azatov:2018kzb,Huang:2018nnq, Angelescu:2018tyl, DaRold:2018moy,Balaji:2018zna, Bansal:2018nwp,Mandal:2018kau,Iguro:2018vqb, Fornal:2018dqn, Kim:2018oih, deMedeirosVarzielas:2019lgb, Zhang:2019hth, Aydemir:2019ynb, deMedeirosVarzielas:2019okf, Cornella:2019hct,Datta:2019tuj, Popov:2019tyc, Bigaran:2019bqv, Hati:2019ufv, Coy:2019rfr, Balaji:2019kwe, Crivellin:2019dwb, Cata:2019wbu,Altmannshofer:2020axr,Cheung:2020sbq,Saad:2020ucl,Saad:2020ihm,Dev:2020qet,Crivellin:2020ukd,Crivellin:2020tsz,Gherardi:2020qhc,Babu:2020hun,Bordone:2020lnb,Crivellin:2020mjs,Crivellin:2020oup,Hati:2020cyn,Dorsner:2021chv,Angelescu:2021lln,Marzocca:2021azj,Crivellin:2021egp,Perez:2021ddi,Crivellin:2021ejk,Zhang:2021dgl,Bordone:2021usz,Carvunis:2021dss,Marzocca:2021miv,BhupalDev:2021ipu,Allwicher:2021rtd,Wang:2021uqz,Bandyopadhyay:2021pld,Qian:2021ihf,Fischer:2021sqw,Gherardi:2021pwm,Crivellin:2021lix,London:2021lfn,Bandyopadhyay:2021kue,Husek:2021isa,Afik:2021xmi,Belanger:2021smw,Chowdhury:2022dps, Heeck:2022znj,Julio:2022bue} requires these particles to have masses of order TeV. Remarkably, these TeV scale scalar leptoquarks (SLQs) can also participate in neutrino mass generation.

In this work, we hypothesize that neutrino masses and LFUV have a common new physics origin. Motivated by this unified framework, we propose a new radiative neutrino mass generation model where scalar leptoquarks, at the leading order, induce tiny neutrino masses as two–-loop quantum corrections. If these LQs reside close to the TeV scale, in addition to incorporating neutrino oscillation data, the proposed model has the potential to address the flavor anomalies mentioned above. However, addressing flavor anomalies demands some of the Yukawa couplings to be of order unity. Therefore, with the TeV-scale LQs, charged lepton flavor violating (cLFV) processes are inevitable, which are also clear signals of new physics. In what follows, we first provide the details of our model and derive the neutrino mass matrix in a general gauge.  From the derived neutrino mass formula, we carry out a comprehensive phenomenological study of the neutrino sector as well as cLFV, which provides the most stringent constraints on the model parameters. Specifically, we investigate a few minimal benchmark scenarios with a limited number of Yukawa parameters without assuming any strong hierarchy among them. We scrutinize these textures for their ability to satisfy neutrino observables and assess cLFV processes with a detailed numerical study using Markov chain Monte Carlo analysis. Finally, we illustrate how $(g-2)_\mu$, the most prominent flavor anomalies, can be addressed while satisfying all LFV constraints and neutrino oscillation data.

This paper is organized in this way: our newly-proposed model of neutrino mass is introduced in Section~\ref{sec:Proposed}, and the detailed derivation of the neutrino mass formula is given in Section~\ref{sec:Formula}. In Sec.~\ref{sec:LFV}, we work out in details all charged lepton flavor violating processes that occur in this model and present the results in Sec.~\ref{sec:results}. Finally, we give our  conclusion in Section~\ref{sec:Con}.

\section{Proposed model}\label{sec:Proposed}

\begin{figure}[b!]
    \centering
    \includegraphics[scale=0.5]{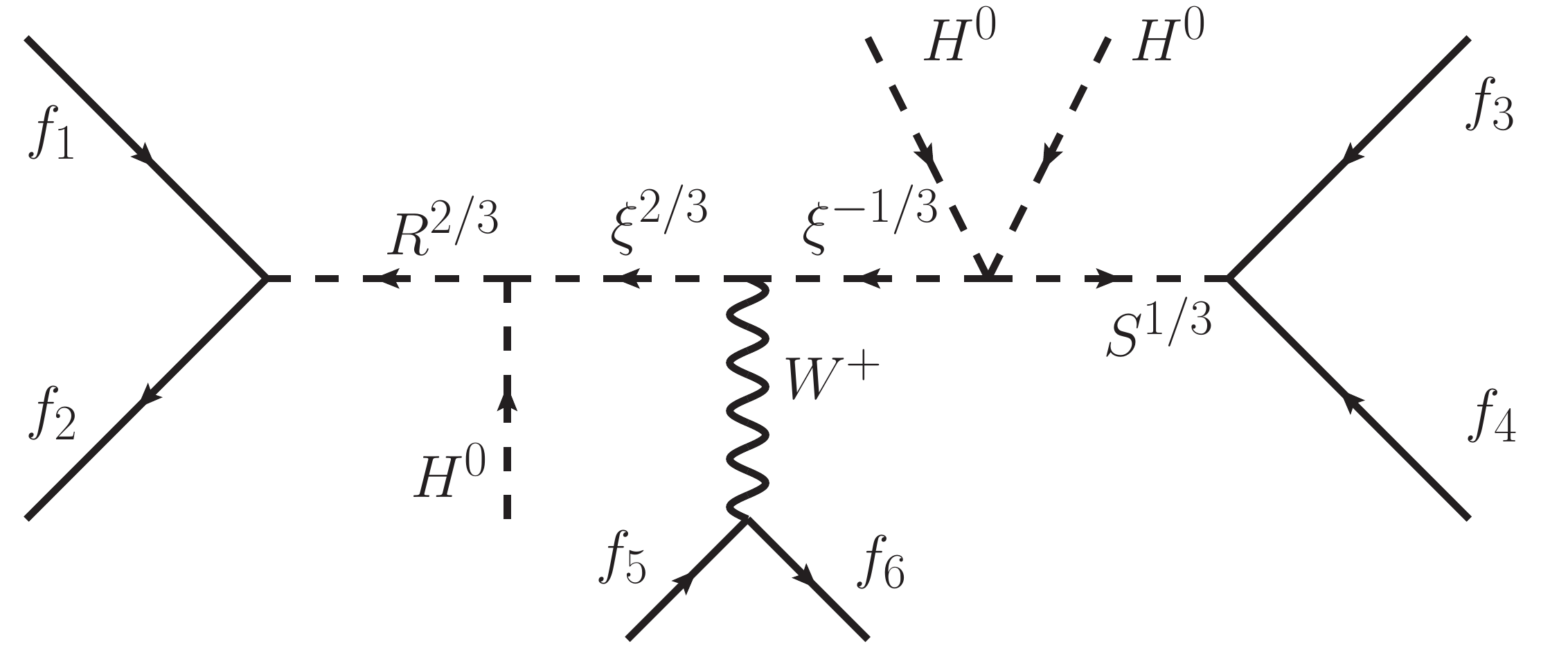}
    \caption{New physics operator leading to non-zero neutrino masses in our proposed model. Here $f_i$ represents SM fermions. From the left-most (right-most) vertex, it is clear that $R_2$ ($S_1$) carries $F=0$ ($F=-2$), where $F=3B+L$ is known as the fermion number. }
    \label{operator}
\end{figure}

The new neutrino mass model proposed in this work consists of the following three BSM scalar multiplets:
\begin{align}
R_2(3,2,7/6)&\equiv R=\left( \begin{array}{c} R^{5/3} \\ R^{2/3} \end{array} \right),
\\
S_1(\overline 3,1,+1/3)
&\equiv S=S^{1/3},
\\
\xi_3(3,3,2/3)&\equiv \xi= \left( \begin{array}{cc} \frac{\xi^{2/3}}{\sqrt{2}} & \xi^{5/3} 
\\ \xi^{-1/3} & -\frac{\xi^{2/3}}{\sqrt{2}} \end{array} \right). 
\end{align}
Numbers in parentheses stand for quantum numbers of each field under $SU(3)_C\times SU(2)_L\times U(1)_Y$ gauge groups.

The SM Higgs is denoted as $H(1,2,1/2)=\left( H^+, H^0 \right)^T$.  This scalar field will get a nonzero vacuum expectation value (vev) $v\equiv \langle H\rangle = 174$ GeV during the spontaneous breaking of  the electroweak (EW) symmetry.  The fermion sector does not change. It contains the same particle content as in the SM
\begin{align}
   &L_i(1,2,-1/2)=\begin{pmatrix}
    \nu_{iL} \\ e_{iL} 
    \end{pmatrix}, \quad
    Q_i(3,2,1/6) = \begin{pmatrix}
    u_{iL} \\ d_{iL}
    \end{pmatrix}, \nonumber \\
    &e_i^c(1,1,1),~~~u_i^c(\bar{3},1,-2/3),~~~ d_i^c(\bar{3},1,+1/3),
\end{align}
where $i$ indicates generation index and $\psi^c\equiv C\bar{\psi_R}^T$ denotes the charge conjugate of the right-handed field.

Among three new scalars introduced, only $R$ and $S$ can be considered as LQs. They interact with the SM fermions through the following Yukawa interactions
\begin{align}
\mathcal{L}_Y^{\rm new} =&  f^L_{ij} u^c_i R \cdot L_j + f^R_{ij} R^\dagger Q_i e^c_j
+ y^L_{ij} S Q_i \cdot L_j + y^R_{ij} u^c_i e^c_j S^\dagger + {\text h.c.}
\label{new-yuk}
\end{align}
In order to avoid unnecessary cluttered notation, we have used ``$\cdot$'' to denote an SU(2) contraction, e.g., $L\cdot Q \equiv L^aQ^b\epsilon_{ab}$ with $\epsilon$ being the antisymmetric tensor ($\epsilon_{12}=-\epsilon^{12}=1$) and $a,b=1,2$ being SU(2) indices. 

All terms in \eq{new-yuk} conserve both baryon ($B$) and lepton ($L$) numbers. This can be seen, for instance, by setting $(B,L)$  to $(1/3,-1)$ and $(-1/3,-1)$ for $R$ and $S$ fields, respectively. Such assignments forbid diquark terms,  $QQS^\dagger$ and $u^cd^cS$, despite being allowed by the gauge symmetry. Note that it is important to have a globally conserved $B$, or else a rapid proton decay will take place in our theory. On the contrary, lepton number $L$ is broken, as required to generate non-zero neutrino mass, by the following non-trivial terms in the scalar potential:
\begin{align}
V \supset &\; \lambda S^\dagger H^T\epsilon \xi^\dagger H + \mu R^\dagger \xi H \nonumber \\
= & ~\lambda S^\dagger \left[-\sqrt{2}\xi^{-2/3}H^+ H^0 - \xi^{1/3}H^0 H^0 + \xi^{-5/3}H^+ H^+ \right] \nonumber \\ & + \mu \bigg\{
\frac{1}{\sqrt{2}} R^{-5/3}\xi^{2/3} H^+
+ R^{-5/3}\xi^{5/3}H^0 
+ R^{-2/3}\xi^{-1/3}H^+ -\frac{1}{\sqrt{2}}R^{-2/3}\xi^{2/3}H^0 
\bigg\} + {\text h.c.}
\label{sc-pot}
\end{align}
One should have noticed that terms in Eq.~\eqref{sc-pot} still conserve the baryon number with $\xi$ field carrying opposite (same) baryon number as that of $S$ ($R$) field. In the above equation, the two parameters, i.e., $\lambda$ and $\mu$, can be made real by absorbing their respected phases into scalar fields. 

The $\Delta L=2$ effective operators, depicted in Fig.~\ref{operator}, must contain the product of $\Delta F=0$ and $\Delta F=2$ couplings, i.e., any combination of  $y^{L,R}f^{L,R}$, so there are four kinds of $\Delta L=2$ operators that can be generated within this model after integrating out heavy scalar states. It is required that such operators involve gauge bosons, or else  they will vanish by $SU(2)$ symmetry. The four operators will contain $(H D^\mu H)$ multiplied by the following combinations: 
\begin{center}
\begin{tabular}{ll}
    (i) & $(u^c H D_\mu L)(LQ)$, \\[0.2em]
    (ii) & $(u^c H D_\mu L)\bar{e^c}\bar{u^c}$, \\[0.2em]
    (iii) & $(\bar{e^c} D_\mu \bar Q H)(LQ)$, \\[0.2em]
    (iv) &  $(\bar{e^c} H D_\mu \bar Q)\bar{u^c}\bar{e^c}$.
\end{tabular}
\end{center}
Note that the $SU(2)$ contraction occurs on fields inside parentheses. In addition, the covariant derivative can also act on other fields. All but the operator (iv) will lead to neutrino masses at two-loop level.

Scalar terms in \eq{sc-pot} will cause  mixing among $\xi^{1/3}-S^{1/3}$, $\xi^{2/3}-R^{2/3}$, and $\xi^{5/3}-R^{5/3}$ components. The mass matrices of leptoquarks relevant for neutrino mass generation are 
\begin{align}
    &M^2_{\chi^{2/3}} = \begin{pmatrix}
    m_\xi^2 & -\lambda v^2 \\ -\lambda v^2 & m_R^2
    \end{pmatrix},
\nonumber \\
    & M^2_{\chi^{1/3}} = \begin{pmatrix}
    m_\xi^2 & -\mu v/\sqrt{2} \\ -\mu v/\sqrt{2} & m_S^2
    \end{pmatrix},
\end{align}
where $m_R$ and $m_\xi$ are the bare masses of $R$ and $\xi$, respectively.
These two mass matrices can be diagonalized by performing the following rotations
\begin{align}
    \text{diag.}\left( M_1^2,M_2^2\right)
    &= U_\phi M^2_{\chi^{2/3}} U^T_\phi, \\
    \text{diag.}\left( M_3^2,M_4^2\right)
    &= U_\theta M^2_{\chi^{1/3}} U^T_\theta,
\end{align}
where
\begin{align}
    U_x = \begin{pmatrix}
    c_x & s_x \\ -s_x & c_x
    \end{pmatrix}.
\end{align}
Here $c_x,s_x$ stand for $\cos x,\sin x$. In terms of scalar mass parameters, the two mixing angles are given by
\begin{align}
\tan2\theta = \frac{-2\lambda v^2}{m_\xi^2-m_S^2}, \quad \tan2\phi = \frac{-\sqrt{2}\mu v}{m_\xi^2-m_R^2}.
\label{mixing-angles}
\end{align}
Furthermore, the mass eigenvalues of $\chi^{2/3}_{1,2}$ and $\chi^{1/3}_{1,2}$ are found to be
\begin{align}
&M_{1,2}^2 = \frac{1}{2}\left[m_\xi^2+m_R^2 \pm \sqrt{(m_\xi^2-m_R^2)^2 + 2\mu^2v^2} \right], \\
&M_{3,4}^2 = \frac{1}{2}\left[m_\xi^2+m_S^2 \pm \sqrt{(m_\xi^2-m_S^2)^2 + 4\lambda^2v^4} \right].
\end{align}
Note that $M^2_{1,2}$ and $M^2_{3,4}$ can be the larger or the smaller of the two mass eigenvalues. They are defined such that
\begin{align}
    M_1^2\cos^2\phi + M_2\sin^2\phi = M_3^2\cos^2\theta + M_4 \sin^2\theta.
    \label{angle-relation}
\end{align}

In terms of mass eigenvalues, we come out with an alternative way of writing Eq. \eqref{mixing-angles}, namely
\begin{align}
\sin2\theta = \frac{-2\lambda v^2}{M_3^2-M_4^2} \quad {\rm and} \quad \sin2\phi = \frac{-\sqrt{2}\mu v}{M_1^2-M_2^2},
\label{mix-mass}
\end{align}
from which both $\lambda$ and $\mu$ can be written in terms of mass eigenvalues
\begin{align}
&\sqrt{2}\lambda v = \frac{-(M_3^2-M_4^2)}{v} \sin2\theta,  
\\&
\mu = \frac{-(M_1^2-M_2^2)}{v}\sin2\phi.   
\end{align}

Having rotated the scalars into their mass eigenstates, we now do the same for Yukawa interactions of \eq{new-yuk}. Without loss of generality, we can define all couplings in Eq.~\eqref{new-yuk} in the  charged lepton mass diagonal basis. If we assume further that the up-type quarks be diagonal as well, \eq{new-yuk} becomes
\begin{align}
\mathcal{L}\supset &~ f^L_{ij} u^c_i \left[ (U_\psi)_{a2} \chi^{5/3}_a e_{Lj} - (U_\phi)_{a2} \chi^{2/3}_a \nu_{Lj} \right]  
\nonumber \\
& +f^R_{ij} \left[(U_\psi)_{a2} \chi^{-5/3}_a u_{Li} + (U_\phi)_{a2} \chi^{-2/3}_a V_{ik} d_{Lk}  \right] e^c_j \nonumber \\
& + y^L_{ij} (U_\theta)_{a2} \left(  u_{Li}e_{Lj}- V_{ik}d_{Lk}\nu_{Lj} \right) \chi^{1/3}_a \nonumber \\
& + y^R_{ij} (U_\theta)_{a2} u^c_i e^c_j \chi^{-1/3}_a,
\label{yuk-mass}
\end{align}
where $V$ is the Cabibbo-Kobayashi-Maskawa mixing matrix.  Note that the results are not affected by changing the up-type to the down-type diagonal basis. Since two choices of the \textit{Yukawa coupling textures}, namely, ``up-type'' and ``down-type'' mass-diagonal basis are widely used in the literature, in the following text, we provide the neutrino mass formula in both these scenarios.  

As one can see from Fig.~\ref{operator}, the neutrino mass generation requires interaction between LQs and $W$ boson, originating from the $SU(2)$ covariant derivatives
\begin{align}
D_\mu R = \left(\partial_\mu -i\frac{g}{2}W^a_\mu\sigma^a -i\frac{7}{6}g'B_\mu\right)R, 
\nonumber
\\
D_\mu \xi = \partial_\mu\xi- i\frac{g}{2}\left[W_\mu^a\sigma^a,\xi\right]-i\frac{2g'}{3}B_\mu,
\label{cov-der}
\end{align}
where $g,g'$ are the $SU(2)$ and $U(1)_Y$ gauge couplings and $\sigma^a$ are Pauli matrices. Based from Eq. \eqref{cov-der},   the $\xi^{2/3}$--$\xi^{1/3}$--$W$  vertex can be derived from the triplet kinetic term, that is,
\begin{align}
   &(D_\mu \xi)^\dagger (D^\mu\xi) 
   \nonumber\\
   & \to ig\left[\xi^{-1/3}\partial_\mu\xi^{-2/3}-(\partial_\mu\xi^{-1/3})\xi^{-2/3}\right]W^{+\mu}.
\end{align}
After rotating the corresponding fields to their mass eigenstates, we obtain
\begin{align}
    \mathcal{L}^{\rm cc}_{\rm scalar} = &~~ ig (U_\theta)_{a1}(U_\phi)_{b1} \nonumber \\
    & \times \left[(\partial_\mu\chi^{-1/3}_a)\chi^{-2/3}_b - \chi^{-1/3}_a(\partial_\mu \chi^{-2/3}_b)\right]W^{+\mu}.
    \label{sc-cc}
\end{align}
In this model, we work in the general $R_\xi$ gauge, so we need to know the LQ interactions with the Goldstone boson. By using Eqs.~\eqref{sc-pot} and \eqref{mix-mass}, the LQs--Goldstone interactions are found to be 
\begin{align}
\mathcal{L} &\supset 
    g \left(\frac{M_b^2-M_{a+2}^2}{m_W} \right) (U_\theta)_{a1}(U_\phi)_{b1} \chi^{-1/3}_a\chi^{-2/3}_bH^+ + {\rm h.c.}.
    \label{sc-goldstone}
\end{align}

\section{Neutrino mass formula}\label{sec:Formula}
\begin{figure}[t!]
    \centering    
    \includegraphics[scale=0.5]{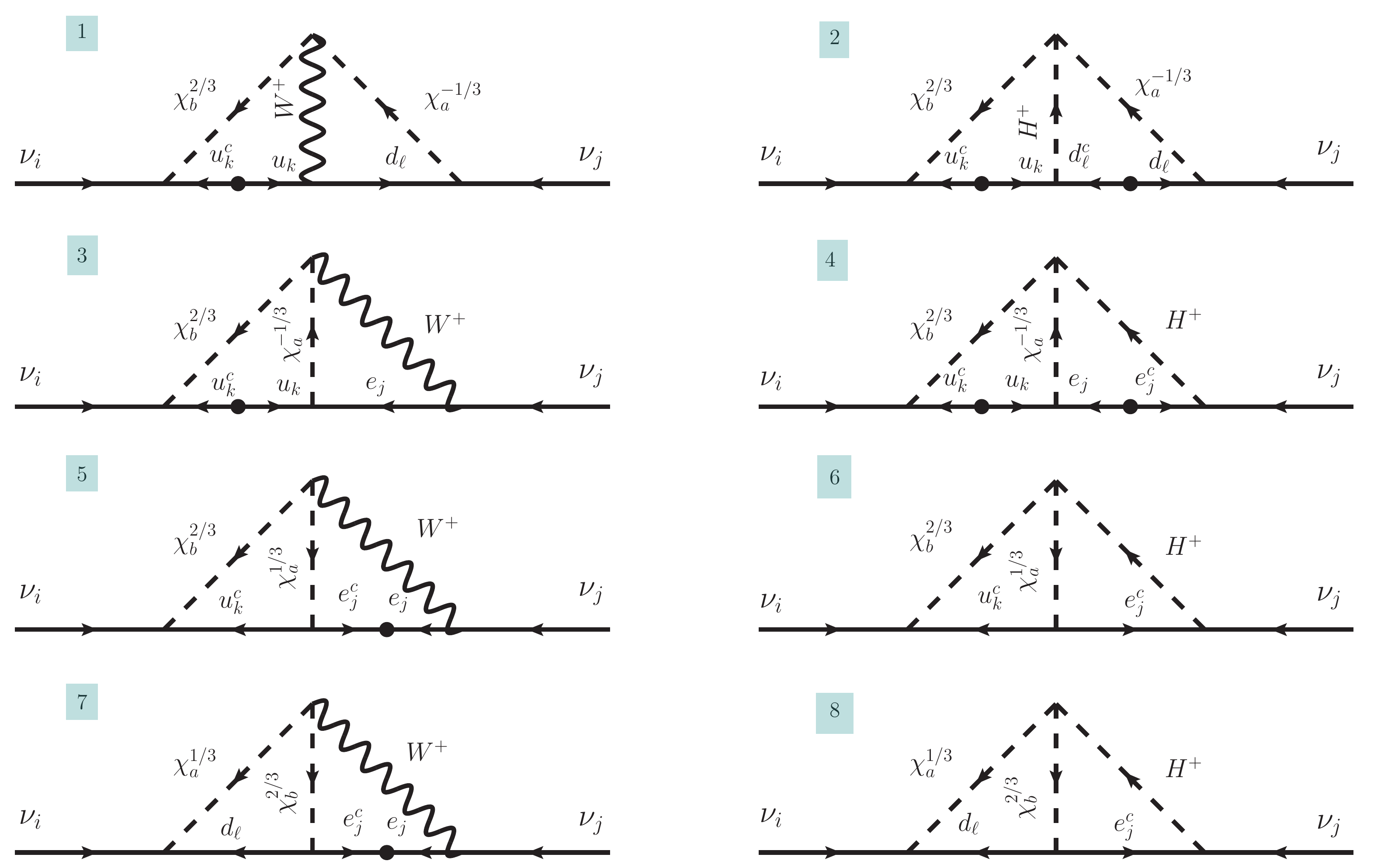}
    \caption{All two--loop diagrams leading to neutrino mass generation. These Feynman diagrams are presented in the mass eigenstate basis.}
    \label{Feynman}
\end{figure}

Armed with all interactions given in Eqs.~\eqref{yuk-mass}, \eqref{sc-cc}, and \eqref{sc-goldstone}, we are ready to construct diagrams leading to neutrino masses. Since there are three different coupling products, there will be three subgroups contributing to neutrino masses. Each contribution is presented in Fig.~\ref{Feynman}. Since neutrino masses are Majorana in nature, in addition to the diagrams shown, there is  another set of diagrams with internal particles replaced by their charge conjugates.  The sum of the two sets of diagrams will result in the neutrino mass matrix being symmetric. 

As mentioned before, in evaluating the neutrino mass diagrams, we  work in the $R_\xi$ gauge.  Therefore, the neutrino mass diagrams will contain gauge parameter $\xi$ dependent terms (not to be confused with $\xi$ multiplet). Such terms will later disappear after we sum over all diagrams. The resulted neutrino mass matrix in the up-type quark diagonal basis  can be written as

\begin{align}
    (\mathcal{M}_\nu)_{ji} =  \frac{3g^2 m_t}{\sqrt 2(16\pi^2)^2} & \left\{ 
 \left[ y^L_{mj}V_{ml}V^{\ast}_{kl}(D_u)_k f^L_{ki} + f^L_{kj} (D_u)_k V^{\ast}_{kl} V_{ml} y^L_{mi} \right] \hat I_{jkl}    
 \right. \nonumber \\
 & \left. 
 +  \frac{m_\tau}{m_t} (D_\ell)_j \left[  y^{R\ast}_{kj}f^L_{ki} + f^L_{kj} y^{R*}_{ki} \right] \tilde I_{jk}
  \right. \nonumber \\
 & \left.
 + \frac{m_\tau}{m_t} (D_\ell)_j \left[ f^{R\ast}_{kj} V^{\ast}_{kl}V_{ml} y^L_{mi} + y^L_{mj} V_{ml}V^{\ast}_{kl}f^{R\ast}_{ki} \right]  \bar I_{jl} 
\right\}. 
\label{nu-mass-up}  
\end{align}

Here the factor of 3 accounts for the exchange of color states inside the loops, whereas $D_u$ and $D_\ell$ are the normalized mass matrices of up-type quarks and charged leptons, respectively
\begin{align}
D_u = \text{diag.}\left(\frac{m_u}{m_t}, \frac{m_c}{m_t}, 1 \right),~~
D_\ell = \text{diag.}\left(\frac{m_e}{m_\tau}, \frac{m_\mu}{m_\tau}, 1 \right).
\end{align}

Similarly, in the basis where down-type quark mass matrix is diagonal, we have
\begin{align}
    (\mathcal{M}_\nu)_{ji} = & \frac{3g^2 m_t}{\sqrt 2(16\pi^2)^2} \left\{ 
 \left[ y^L_{lj}V^{\ast}_{kl}(D_u)_k f^L_{ki} + f^L_{kj} (D_u)_k V^{\ast}_{kl} y^L_{li} \right] \hat I_{jkl} \right. \nonumber \\
 &   +  \frac{m_\tau}{m_t} (D_\ell)_j \left[  y^{R\ast}_{kj}f^L_{ki} + f^L_{kj} y^{R*}_{ki} \right] \tilde I_{jk} \nonumber \\
 & \left. + \frac{m_\tau}{m_t} (D_\ell)_j \left[ f^{R\ast}_{lj} y^L_{li} + y^L_{lj} f^{R\ast}_{li} \right]  \bar I_{jl} 
\right\}. 
\label{nu-mass-down}
\end{align}
One should note that  Eqs.~\eqref{nu-mass-up} and \eqref{nu-mass-down} are equivalent, as one can recover the latter, for instance, by redefining $y^L  \to V^{\ast} y^L$ and $f^R \to V^{\ast}f^R$.  This reflects the basis independence mentioned previously. 

The loop integrals shown in Eqs.~\eqref{nu-mass-up} and \eqref{nu-mass-down}, i.e., $\hat I_{jkl}, \tilde I_{jk}$, and $\bar I_{jl}$, indicate the contribution of each subgroup. Each of them is defined as
\begin{align}
    \hat I_{jkl} = & (16\pi^2)^2\left[I^{(1)}_{kl} + I^{(2)}_{kl} + I^{(3)}_{jk} + I^{(4)}_{jk} \right], \nonumber \\
    \tilde I_{jk} = & (16\pi^2)^2 \left[I^{(5)}_{jk} + I^{(6)}_{jk} \right], \\
    \bar I_{jl} =& (16\pi^2)^2 \left[I^{(7)}_{jl} + I^{(8)}_{jl} \right], \nonumber
\end{align}
with  $I^{(n)}_{ij}$ denoting the dimensionless loop function for the $n$-th diagram, that is,

\begin{align}
I^{(1)}_{kl} =~& 
-(U_\theta)_{a1}(U_\theta)_{a2}(U_\phi)_{b1}(U_\phi)_{b2} \int \frac{d^4k}{(2\pi)^4}  \int \frac{d^4q}{(2\pi)^4} \frac{1}{k^2-m^2_W}  \nonumber \\
& \times \frac{1}{(q+k)^2-m^2_{u_k}} 
\frac{1}{(q+k)^2-M^2_{b}} \frac{1}{q^2-m^2_{d_l}}\frac{1}{q^2-M^2_{a+2}} \nonumber \\
& \times\left[
\slashed q(2\slashed q+\slashed k)+ \frac{\slashed q \slashed k}{k^2} \left( -1+\xi \frac{k^2-m^2_W}{k^2-\xi m^2_W} \right) k\cdot(2q+k)
\right]\;,
\label{eq:int1}
\\
I^{(2)}_{kl} =~ & (U_\theta)_{a1}(U_\theta)_{a2}(U_\phi)_{b1}(U_\phi)_{b2}  \int \frac{d^4k}{(2\pi)^4}  \int \frac{d^4q}{(2\pi)^4}
\frac{1}{k^2} \left( \frac{M_b^2-M_{a+2}^2 }{m^2_W} \right)
\left(  1+\xi \frac{m^2_W}{k^2-\xi m^2_W}  \right) \nonumber \\
& \times \left(1+\frac{\slashed q \slashed k}{q^2-m_{d_l}^2} \right) \frac{1}{(q+k)^2-m^2_{u_k}} \frac{1}{q^2-M^2_{a+2}}\frac{1}{(q+k)^2-M^2_{b}}\;,
\label{eq:int2}
\\
I^{(3)}_{jk} =~ & -(U_\theta)_{a1}(U_\theta)_{a2}(U_\phi)_{b1}(U_\phi)_{b2}  \int \frac{d^4k}{(2\pi)^4}  \int \frac{d^4q}{(2\pi)^4}
\frac{1}{k^2}\left(1+\frac{m^2_{e_j}}{k^2-m^2_{e_j}}  \right)  \nonumber \\
& \times  \frac{1}{k^2-m^2_W} \frac{1}{q^2-M^2_{a+2}}\frac{1}{(q+k)^2-M^2_{b}} \frac{1}{(q+k)^2-m^2_{u_k}} \nonumber \\
& \times \left[
(2\slashed q+\slashed k)\slashed k +  \left( -1+\xi \frac{k^2-m^2_W}{k^2-\xi m^2_W} \right) k\cdot(2q+k)
\right]\;,
\label{eq:int3}
\\
I^{(4)}_{jk} =~ & (U_\theta)_{a1}(U_\theta)_{a2}(U_\phi)_{b1}(U_\phi)_{b2}  \int \frac{d^4k}{(2\pi)^4}  \int \frac{d^4q}{(2\pi)^4}  \frac{1}{k^2}
\left(\frac{m^2_{e_j}}{k^2-m^2_{e_j}} \right)  \left(\frac{M_b^2-M_{a+2}^2 }{m^2_W} \right) \nonumber \\
& \times \left(  1+\xi \frac{m^2_W}{k^2-\xi m^2_W}  \right) \frac{1}{q^2-M^2_{a+2}}\frac{1}{(q+k)^2-M^2_{b}} \frac{1}{(q+k)^2-m^2_{u_k}}\;,
 \label{eq:int4}
 \\
I^{(5)}_{jk}  =~ & -(U_\theta)_{a1}(U_\theta)_{a2}(U_\phi)_{b1}(U_\phi)_{b2}  \int \frac{d^4k}{(2\pi)^4}  \int \frac{d^4q}{(2\pi)^4} \nonumber \\
& \times \frac{1}{k^2-m_W^2} \frac{1}{k^2-m_{e_j}^2} \frac{1}{q^2-M_{a+2}^2} \frac{1}{(q+k)^2-M_b^2} \frac{1}{(q+k)^2-m_{u_k}^2} \nonumber \\
& \times \left[
(2\slashed q + \slashed k)(\slashed q + \slashed k)  + \frac{\slashed k (\slashed q+\slashed k)}{k^2}\left(-1 + \xi \frac{k^2-m_W^2}{k^2-\xi m_W^2} \right) k\cdot(2q+k) \right]\;, 
\label{eq:int5}
\\
I^{(6)}_{jk}  =~ & (U_\theta)_{a1}(U_\theta)_{a2}(U_\phi)_{b1}(U_\phi)_{b2}  \int \frac{d^4k}{(2\pi)^4}  \int \frac{d^4q}{(2\pi)^4} 
\frac{\slashed k (\slashed q + \slashed k)}{k^2}  \left(\frac{M_b^2-M_{a+2}^2}{m_W^2}  \right) \nonumber \\
& \times \left( 1 + \xi \frac{m_W^2}{k^2-\xi m_W^2} \right)  \frac{1}{k^2-m_{e_j}^2}  \frac{1}{q^2-M_{a+2}^2} \frac{1}{(q+k)^2-M_b^2}\frac{1}{(q+k)^2-m_{u_k}^2}\;,
\label{eq:int6}
\\
I^{(7)}_{jl} =~ & (U_\theta)_{a1}(U_\theta)_{a2}(U_\phi)_{b1}(U_\phi)_{b2}  \int \frac{d^4k}{(2\pi)^4}  \int \frac{d^4q}{(2\pi)^4}  \frac{1}{k^2-m_{W}^2} \frac{1}{k^2-m_{e_j}^2} \nonumber \\
& \times \frac{1}{(q+k)^2-m_{d_l}^2} \frac{1}{(q+k)^2-M_{a+2}^2} \frac{1}{q^2-M_b^2} \nonumber \\
& \times \left[
(2\slashed q + \slashed k)(\slashed q + \slashed k)  + \frac{\slashed k (\slashed q + \slashed k)}{k^2}\left(-1 + \xi \frac{k^2-m_W^2}{k^2-\xi m_W^2} \right) k\cdot(2q+k) \right]\;,
\label{eq:int7}
\\
I^{(8)}_{jl}  =~ & (U_\theta)_{a1}(U_\theta)_{a2}(U_\phi)_{b1}(U_\phi)_{b2}  \int \frac{d^4k}{(2\pi)^4}  \int \frac{d^4q}{(2\pi)^4} 
\frac{\slashed k  (\slashed q + \slashed k)}{k^2} \left(\frac{M_b^2-M_{a+2}^2}{m_W^2} \right) \nonumber \\
& \times \left( 1 + \xi \frac{m_W^2}{k^2-\xi m_W^2} \right)  \frac{1}{k^2-m_{e_j}^2} \frac{1}{(q+k)^2-m_{d_l}^2} \frac{1}{(q+k)^2-M_{a+2}^2} \frac{1}{q^2-M_b^2}\;.
\label{eq:int8}
\end{align}

The cancellation of terms containing the gauge parameter $\xi$ can be inferred directly from Eqs.~\eqref{eq:int1}-\eqref{eq:int8}. To see how  this cancellation takes place, it is desirable to express $k\cdot(2q+k)$, which appears in all $W$-mediated diagrams, as
\begin{align}
 k \cdot (2q+k) = & \left[(q+k)^2-M_b^2-q^2+M_{a+2}^2\right]  + M_b^2-M_{a+2}^2. 
\end{align}
Terms inside parentheses will cancel the LQ propagators, particularly those appearing in diagrams 1, 3, and 5. Thus, they will vanish by the  orthogonality of LQ mixing matrices. The remaining terms, which are proportional to $M_b^2-M_{a+2}^2$, will make such loop terms have the same coefficients but opposite signs with  the corresponding Goldstone loop integrals, allowing the cancellation of $\xi$-dependent terms. For diagram 7, due to momentum switch between $\chi^{1/3}$ and $\chi^{2/3}$ (see Fig.~\ref{Feynman}), we have instead $k\cdot (2q+k) \to M_{a+2}^2-M_b^2$. The cancellation of $\xi$-dependent terms in this case too can be foreseen right away. The gauge parameter cancellation indicates further that all two-loop diagrams presented in Fig.~\ref{Feynman} are the complete set of diagrams generating neutrino masses at the lowest order.

We are, then, left with gauge-independent terms. It is straightforward to evaluate the integrals, from which we get

\begin{align}
   \hat I_{jkl} =~&  -\frac{1}{4} \sin2\theta\sin2\phi \sum_{a,b=1}^2 (-1)^{a+b} \frac{1}{(M_b^2-m_{u_k}^2)(M_{a+2}^2-m_{d_l}^2)} 
   \int_0^1 dx \int_0^\infty dt \frac{t}{t+m_W^2} \nonumber \\ 
   & \times \left\{ xt \left[6x-5 + \left(\frac{M_b^2-M_{a+2}^2}{m_W^2}\right) \right] \left[ \ln \frac{\Delta(x,t;M_b,M_{a+2})}{\Delta(x,t;M_b,m_{d_l})} - \ln \frac{\Delta(x,t;m_{u_k},M_{a+2})}{\Delta(x,t;m_{u_k},m_{d_l})}\right] \right. \nonumber \\
   & \quad -4 \left[A(x;M_b,M_{a+2}) \ln \frac{\Delta(x,t;M_b,M_{a+2})}{m_W^2} + A(x;m_{u_k},m_{d_l}) \ln \frac{\Delta(x,t;m_{u_k},m_{d_l})}{m_W^2} \right. \nonumber \\
   & \quad \left. -A(x;M_b,m_{d_l}) \ln \frac{\Delta(x,t;M_b,m_{d_l})}{m_W^2} - A(x;m_{u_k},M_{a+2}) \ln \frac{\Delta(x,t;m_{u_k},M_{a+2})}{m_W^2}\right] \nonumber \\ 
   & \quad + \left(\frac{M_{a+2}^2-m_{d_l}^2}{t+m_{e_j}^2}\right) \left[ (2x-1) + \left( \frac{M_b^2-M_{a+2}^2}{m_W^2} \right) \right] t
   \ln \frac{\Delta(x,t;M_b,M_{a+2})}{\Delta(x,t;m_{u_k},M_{a+2})} \bigg\},
  \\
   \tilde I_{jk}  =~&  \frac{1}{4} \sin2\theta\sin2\phi \sum_{a,b=1}^2 (-1)^{a+b}\frac{1}{M_b^2-m_{u_k}^2} 
   \int_0^1 dx \int_0^\infty dt \frac{t}{(t+m_W^2)(t+m_{e_j}^2)} \nonumber \\
   & \times \left\{\left[ 1-6x - \left(\frac{M_b^2-M_{a+2}^2}{m_W^2} \right) \right] (1-x) t  \ln \frac{\Delta(x,t;M_b,M_{a+2})}{\Delta(x,t;m_{u_k},M_{a+2})} \right. \nonumber \\
   & \left. -4A(x;M_b,M_{a+2}) \ln \frac{\Delta(x,t;M_b,M_{a+2})}{m_W^2} + 4A(x;m_{u_k},M_{a+2}) \ln \frac{\Delta(x,t;m_{u_k},M_{a+2})}{m_W^2} \right\}, \\
   \bar I_{jl}  =~&  -\frac{1}{4} \sin2\theta\sin2\phi \sum_{a,b=1}^2 (-1)^{a+b}\frac{1}{M_{a+2}^2-m_{d_l}^2} 
   \int_0^1 dx \int_0^\infty dt \frac{t}{(t+m_W^2)(t+m_{e_j}^2)} \nonumber \\
   & \times \left\{\left[ 1-6x - \left(\frac{M_{a+2}^2-M_{b}^2}{m_W^2} \right) \right] (1-x) t  \ln \frac{\Delta(x,t;M_{a+2},M_b)}{\Delta(x,t;m_{d_l},M_b)} \right. \nonumber \\
   & \left. -4A(x;M_{a+2},M_b) \ln \frac{\Delta(x,t;M_{a+2},M_b)}{m_W^2} + 4A(x;m_{d_l},M_b) \ln \frac{\Delta(x,t;m_{d_l},M_b)}{m_W^2} \right\},
\end{align}
where only terms relevant to neutrino masses are kept. 
In those loop integral expressions, we have introduced a parameter 
\begin{align}
    \Delta(x,t;m,M) \equiv x(1-x)t + A(x;m,M),
\end{align} 
with $A(x;m,M) \equiv xm^2 + (1-x)M^2$. 

\begin{figure}[t!]
    \centering
    \includegraphics[width=9cm]{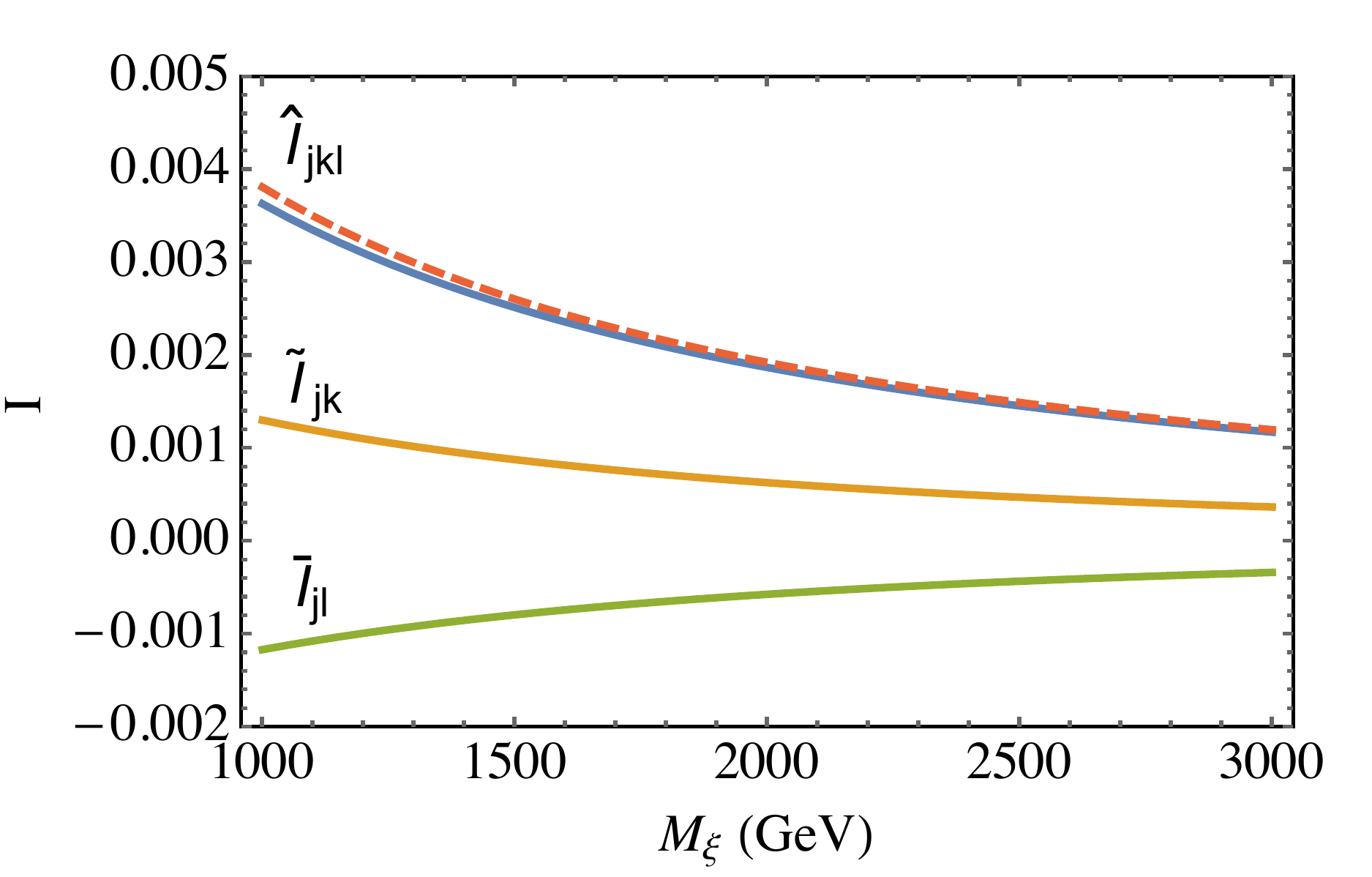}
    \caption{The plot of loop integral of each diagram. The solid line represents the case with the top quark inside the loop, while the dashed line represents the case with the charm quark. It is obvious that the loop integrals depend mildly on fermion masses. Here we use $M_S=1200$ GeV, $M_R=1501$ GeV, $\mu=10$ GeV, and $\lambda=1$.}
    \label{fig:int-plot}
\end{figure}

All loop integrals are finite, and thus can be calculated numerically. In addition, the contributions of light fermion masses are negligible. Therefore, they can be simply omitted from the integrals, which is demonstrated in  Fig.~\ref{fig:int-plot}.

\section{Lepton flavor violation}\label{sec:LFV}
All couplings presented in Eq.~\eqref{new-yuk} naturally generate lepton-flavor violating (LFV) processes, which are strongly constrained. In this section, we will use those constraints to scrutinize our model. For the sake of compactness, we write the Eq.~\eqref{new-yuk} as
\begin{align}
{\cal L} = \bar\ell_i \left(\lambda_L^{ij} P_{R} + \lambda_R^{ij} P_L \right) q_j \phi^\ast + {\rm h.c.}.
\end{align}
In this notation, we define $\ell$ as the charged leptons with $Q_\ell=-1$ and subscripts $L,R$ on $\lambda$'s indicate the chirality of such fields. It is then straightforward to see that there are three types of $\lambda$'s within this model. The mapping of $\lambda_{L,R}$ into couplings given in Eq.~\eqref{new-yuk}  is shown in Table~\ref{tab:conversion}.
\begin{table}[t!]
    \centering
   {\renewcommand{\arraystretch}{1.3}
    \begin{tabular}{|c|c|c|}
    \hline\hline
    \multirow{2}{*}{$q,\phi$}  & \multicolumn{2}{c|}{$\lambda_L,\lambda_R $}  \\ \cline{2-3}
    & Up-mass diagonal & Down-mass diagonal \\
   \hline
    $u,R^{5/3}$ & $(f^L)^\dagger,(f^{R})^T$  & $(f)^{L})^{\dagger},(f^{R})^T V^\dagger$ \\
    \hline
    $u^c,S^{1/3}$ & $(y^{L})^{\dagger},(y^R)^T$ & $(y^{L})^{\dagger}V^T,(y^{R})^T$ \\ \hline 
    $d,R^{2/3}$ & $0,(f^R)^T$ & $0,(f^R)^TV$\\ \hline\hline
    \end{tabular}}
    \caption{Mapping of Yukawa couplings of Eq.~\eqref{new-yuk} into $\lambda_{L,R}$. Note that, in this table, $u^c$ is defined as charge conjugate of up-type quarks, namely $u^c=C\bar u^T$.}
    \label{tab:conversion}
\end{table}

\subsection{$\ell_i \to \ell_k+\gamma$ decay}
Due to flavor-violating nature of Eq.~\eqref{new-yuk}, lepton flavor violating processes in general are expected to occur within this model. The first process we consider is $\ell_i\to\ell_k+\gamma^\ast$ transition, whose effective Lagrangian
\begin{align}
\mathcal{L}_{\ell_i \to \ell_k + \gamma^*} = &~ -m_{\ell_i} \bar{\ell}_k \sigma^{\alpha\beta}\left(\tfrac{e}{2}A_{2L}^\ast P_L + \tfrac{e}{2}A_{2R}^\ast P_R \right)\ell_i F_{\alpha\beta}  \nonumber \\
&~ -\bar\ell_k \gamma^\alpha\left(eA_{1L}^\ast P_L + eA_{1R}^\ast P_R \right)\ell_i A^\beta \left(q^2g_{\alpha\beta} - q_\alpha q_\beta\right) +  {\rm h.c.},
\label{eq:muegamma}
\end{align}
where $F_{\alpha\beta}=\partial_\alpha A_\beta-\partial_\beta A_\alpha$ is the electromagnetic field tensor and $q= p_{\ell_i}-p_{\ell_k}$ is the photon momentum transfer. At the lowest order, this kind of processes arises through penguin-type diagrams exchanging LQs. It is worth noting that, owing to the Ward-Takahashi identity, only dipole terms survive in a lepton decay into an on-shell photon. Its decay width is given by~\cite{Lavoura:2003xp}
\begin{align}
\Gamma(\ell_i \to \ell_k \gamma) =&~ \frac{\alpha_{em}}{4}m_{\ell_i}^5 \left(|A_{2L}|^2+|A_{2R}|^2 \right),
\end{align}
where $\alpha_{em}=e^2/4\pi$ is the electromagnetic fine-structure constant, $m_{\ell_i}$ is the decaying lepton mass, and
\begin{align}
A_{2L} = &~ \frac{3}{16\pi^2} \sum_{\lambda,\phi} \frac{1}{M_\phi^2} \bigg\{ \left[\lambda^{ij}_R\lambda^{kj\ast}_R + \frac{m_{\ell_k}}{m_{\ell_i}} \lambda^{ij}_L\lambda^{kj\ast}_L\right]\left[Q_qF_1(x_j)+Q_\phi F_2(x_j) \right] \nonumber \\
& \qquad + \frac{m_{q_j}}{m_{\ell_i}} \lambda^{ij}_L\lambda^{kj\ast}_R \left[ Q_q F_3(x_j)+ Q_\phi F_4(x_j) \right] \bigg\}, \nonumber \\
A_{2R} =&~ A_{2L} (L \leftrightarrow R).
\label{eq:A2}
\end{align}
In the above equation, 3 is the color factor and $x_j=m_{q_j}^2/M_\phi^2$. The summation is performed over all possible couplings and leptoquark fields, as given in Table~\ref{tab:conversion}. Quantities $Q_q$ and $Q_\phi$ are the corresponding quark and LQ electric charges, which obey $Q_\ell+Q_\phi=Q_q$ with $Q_\ell=-1$. Functions $F_1(x)$ and $F_3(x)$ are evaluated  from diagrams emitting a photon from the quark line, whereas  $F_2(x)$ and $F_4(x)$ from diagrams emitting a photon from the LQ line. They all are given by
\begin{align}
F_1(x) =&~ \frac{2+3x-6x^2+x^3+6x\ln x}{6(1-x)^4},\nonumber \\
F_2(x) =&~ \frac{1-6x+3x^2+2x^3-6x^2\ln x}{6(1-x)^4},\nonumber \\
F_3(x) =&~ \frac{-3+4x-x^2-2\ln x}{(1-x)^3}, \nonumber \\
F_4(x) =& ~ \frac{1-x^2+2x\ln x}{(1-x)^3}.
\label{eq:dipole-func}
\end{align}
Due to the possibility of simultaneous existence of both $\lambda_L$ and $\lambda_R$ in this model, see Table~\ref{tab:conversion}, we can have a chirality-enhanced process, especially when the top  quark is inside the loop. This will lead to severe constraints on the Yukawa couplings. Current and future rates of this kind of processes are presented in Table~\ref{tab:201}.

\begin{table}[t!]
\begin{center}
\begin{tabular}{|c|c|c|}\hline
$\ell_i\to \ell_k\gamma$ & Present bound & Future sensitivity \\\hline\hline

$\mu\to e \gamma$&$4.2\times 10^{-13}$~\cite{MEG:2016leq} &$6\times 10^{-14}$~\cite{Baldini:2013ke}  \\ \hline
$\tau\to e \gamma$&$3.3\times 10^{-8}$~\cite{BaBar:2009hkt} &$\sim 10^{-9}$~\cite{Aushev:2010bq}  \\ \hline
$\tau\to \mu \gamma$&$4.4\times 10^{-8}$~\cite{BaBar:2009hkt} &$\sim 10^{-9}$~\cite{Aushev:2010bq}  \\ \hline

\end{tabular}
\end{center}
\caption{Current experimental bounds on the $BR(\ell_i\to \ell_k\gamma)$. Future sensitivities are presented on the last column. }
\label{tab:201}
\end{table}

\subsection{Lepton 3-body decay}
This kind of processes also occurs at loop level, consisting of photon- and $Z$-penguin diagrams as well as the box diagrams. For the photon-mediated processes, one just needs to attach the photon leg in Eq.~\eqref{eq:muegamma} with a $\ell^+_l\ell^-_l$ pair. Now, the photon is off shell, both $A_{2L,R}$ and $A_{1L,R}$ contribute to the photon-induced effective Lagrangian, written as \cite{Kuno:1999jp}
\begin{align}
{\cal L}_{\ell_i^- \to \ell_k^-\ell_l^+\ell_l^-}^{\gamma-\rm penguin} =&~ -m_{\ell_i} \bar \ell_k \sigma^{\alpha\beta} \left(\tfrac{e}{2}A_{2L}^\ast P_L + \tfrac{e}{2}A_{2R}^\ast P_R \right)\ell_i F_{\alpha\beta} \nonumber \\
&~ \left[\bar\ell_k \gamma^\alpha \left(e^2A_{1L}^\ast P_L + e^2A_{1R}^\ast P_R \right) \ell_i \right] \left[ \bar\ell_l \gamma_\alpha P_L \ell_l + \bar\ell_l \gamma_\alpha P_R \ell_l \right].
\end{align}
It is also straightforward to evaluate $A_{1L},A_{1R}$, which are given by
\begin{align}
A_{1L} =&~ \frac{3}{16\pi^2} \sum_{\lambda,\phi} \frac{1}{M_\phi^2}  \lambda_L^{ij}\lambda^{kj\ast}_L \left[Q_q G_1(x_j) + Q_\phi G_2(x_j) \right], \nonumber \\
A_{1R} =&~ A_{1L} (L \leftrightarrow R),
\end{align}
with 
\begin{align}
G_1(x) =&~\frac{16-45x+36x^2-7x^3+6(2-3x)\ln x}{36(1-x)^4}, \nonumber \\
G_2(x) =&~ \frac{2-9x+18x^2-11x^3+6x^3\ln x}{36(1-x)^4}.
\end{align}

In addition to the aforementioned photonic diagrams, one can also have $Z$-penguin interactions. They are given by
\begin{align}
{\cal L}_{\ell_i^- \to \ell_k^-\ell_l^+\ell_l^-}^{Z-\rm penguin} = & \left[\bar\ell_k \gamma^\alpha \left(e^2 Z_L^\ast P_L + e^2Z_R^\ast P_R \right) \ell_i \right] \left[ g_L^{(\ell)}\bar\ell_l \gamma_\alpha P_L \ell_l + g_R^{(\ell)}\bar\ell_l \gamma_\alpha P_R \ell_l \right] + {\rm h.c.},
\end{align}
with
\begin{align}
Z_{L} =&~ \frac{3}{16\pi^2} \sum_{\lambda,\phi} \frac{\lambda^{ij}_L\lambda^{kj\ast}_L}{m_Z^2\cos^2\theta_W\sin^2\theta_W} \nonumber \\
& \qquad\qquad \times \left\{ g_R^{(q_j)} C_1(x_j) + g_L^{(q_j)}C_2(x_j) - \left[g^{(\phi)} + g^{(\ell_k)}_L\right] \left[C_1(x_j)+C_2(x_j)\right] \right\}, \nonumber \\
Z_R =&~ Z_L (L\leftrightarrow R).
\label{eq:ZL}
\end{align}
In deriving $Z_{L,R}$, we have neglected terms proportional to $m_{\ell_i}$. 
Here $g^{(f)}_{L,R}=T_{3f_{L,R}}-Q_f\sin^2\theta_W$, with $\theta_W$ being the weak mixing angle, and
\begin{eqnarray}
C_1(x) &=& \frac{-1+x^2+2(-2+x)x\ln x}{4(1-x)^2},\nonumber \\
C_2(x) &=& \frac{x(1-x+\ln x)}{(1-x)^2}.
\end{eqnarray}

Similarly, for the box diagrams, we have
\begin{align}
{\cal L}_{\ell_i^- \to \ell_k^-\ell_l^+\ell_l^-}^{\rm box} = &~~ e^2B_{1L}^\ast [\bar\ell_k \gamma^\alpha P_L \ell_i][\bar\ell_l \gamma_\alpha P_L \ell_l] + e^2B_{2L}^\ast [\bar\ell_k \gamma^\alpha P_L \ell_i][\bar\ell_l \gamma_\alpha P_R \ell_l] \nonumber \\
& + e^2B_{3L}^\ast [\bar\ell_k P_L \ell_i][\bar\ell_l P_L \ell_l] + (L \leftrightarrow R) +  {\rm h.c.}
\end{align}
Note that we do not list box operators in the form of $(S\mp P)\times (S\pm P)$, with $S,P$ indicating scalar and pseudoscalar bilinears. This is because they can always be Fierz reordered into the form of $(V\mp A)\times (V\pm A)$ operator, which is already included. The corresponding Wilson's coefficients are found to be
\begin{align}
e^2B_{1L} =&~ \frac{3}{16\pi^2}\sum_{\lambda,\phi} \frac{1}{M_\phi^2} \lambda^{ij}_L\lambda^{kj\ast}_L|\lambda_L^{kn}|^2 b_1(x_j,x_n),
\nonumber \\
e^2B_{2L} = & \frac{3}{16\pi^2}\sum_{\lambda,\phi} \frac{1}{M_\phi^2} \left[ \lambda_L^{ij}\lambda_L^{kj\ast} |\lambda_R^{kn}|^2 b_1(x_j,x_n) - \tfrac{1}{2} \lambda_L^{ij}\lambda_R^{kj\ast} \lambda_R^{kn}\lambda_L^{kn\ast} b_2(x_j,x_n) \right] ,
\nonumber \\
e^2B_{3L} =& \frac{3}{16\pi^2}\sum_{\lambda,\phi} \frac{1}{M_\phi^2}  \lambda_L^{ij}\lambda_R^{kj\ast} \lambda_L^{kn}\lambda_R^{kn\ast} b_2(x_j,x_n) ,
\nonumber \\
B_{iR} =&~ B_{iL} (L \leftrightarrow R), \quad {i=1,2,3}.
\label{box-loop}
\end{align}
One can see that $B_{2L}$ has two terms, but they do not come from the same set of couplings. The first term, coming from momenta of internal quarks, similar to $B_{1L}$, has $(V-A)\times (V+A)$ type. The second one comes through the internal quark chirality flip, which is then Fierz reordered. That explains why it picks  the factor of $-1/2$. The loop functions $b_1$ and $b_2$ are determined to be
\begin{align}
b_1(x_j,x_n)=& -\tfrac{1}{2}\int dt  \frac{t^2}{(t+1)^2(t+x_j)(t+x_n)}, \nonumber \\
b_2(x_j,x_n) =& \sqrt{x_j x_n}\int dt \frac{t}{(t+1)^2(t+x_j)(t+x_n)}.
\end{align}
The factor of $-1/2$ comes from $\int k^\mu k^\nu=\tfrac{1}{2}g^{\mu\nu}\int k^2$, which is later Wick rotated. It is straightforward to evaluate these integrals, yielding
\begin{align}
&b_1(x,y) = \left\{ \begin{array}{c} \frac{-1+x^2-2x\ln x}{2(1-x)^3}~{\rm for}~y=x \\ \\ 
\frac{-1+x-x\ln x}{2(1-x)^2}~{\rm for}~y=0 \\\\
-\tfrac{1}{2}~{\rm for}~x=y=0 
\end{array}
\right.
\\
&b_2(x,y) = \frac{-2x+2x^2-(1+x)x\ln x}{(1-x)^3}~{\rm  for}~y=x.
\end{align}

Combining all interactions mentioned before, we can write the most general $\ell_i^-\to\ell_k^-\ell_l^+\ell_l^-$ effective Lagrangian, namely
\begin{align}
{\cal L}_{\ell_i^- \to \ell_k^-\ell_l^+\ell_l^+} = &~ -m_{\ell_i} \bar\ell_i \sigma^{\alpha\beta} \left(\tfrac{e}{2}A_{2R}P_L + \tfrac{e}{2}A_{2L}P_R \right) \ell_k F_{\alpha\beta} \nonumber \\
&~ - \left[ g_1 (\bar\ell_i P_R \ell_k)(\bar \ell_l P_R \ell_l)  + g_2 (\bar\ell_i P_L \ell_k)(\bar \ell_l P_L \ell_l)\right. \nonumber \\
&\quad  + g_3 (\bar\ell_i \gamma^\alpha P_R \ell_k)(\bar \ell_l \gamma_\alpha P_R \ell_l)  + g_5 (\bar\ell_i \gamma^\alpha P_L \ell_k)(\bar \ell_l \gamma_\alpha P_L \ell_l) \nonumber \\
&\quad \left. + g_5 (\bar\ell_i \gamma^\alpha P_R \ell_k)(\bar \ell_l \gamma_\alpha P_L \ell_l)  + g_6 (\bar\ell_i \gamma^\alpha P_L \ell_k)(\bar \ell_l \gamma_\alpha P_R \ell_l) \right] + \rm h.c.,
\end{align}
where we have followed the notation of Ref.~\cite{Kuno:1999jp}. The coefficients $g_1,...g_6$ consist of all contributions from photon, $Z$, and box diagrams, which are given by
\begin{align}
&g_1 = -e^2B_{3L}, \quad g_2 = -e^2B_{3R}, \\
&g_3 = -e^2\left(A_{1R} + Z_Rg_R^{(\ell)} + B_{1R} \right), \quad
g_4 = -e^2\left(A_{1L} + Z_Lg_L^{(\ell)} + B_{1L} \right),\nonumber \\
&g_5 = -e^2\left(A_{1R} + Z_Rg_L^{(\ell)} + B_{2R} \right), \quad
g_6 = -e^2\left(A_{1L} + Z_Lg_R^{(\ell)} + B_{2L} \right).
\end{align}
From here we can calculate the decay width \cite{Kuno:1999jp,Abada:2014kba}
\begin{align}
\Gamma (\ell^-_i \to \ell^-_k\ell^+_l\ell^-_l) =&~ \frac{m_{\ell_i}^5}{512\pi^3} \nonumber \\
&\times \bigg[ \frac{1}{12(1+\delta_{kl})}\left(|g_1|^2 + |g_2|^2 \right) + \frac{1+\delta_{kl}}{3}\left( |g_3|^2+|g_4|^2\right) +  \frac{1}{3}\left(|g_5|^2+|g_6|^2\right) \nonumber \\
& \qquad +  \left[\frac{16}{3}\ln \frac{m_{\ell_i}}{m_{\ell_n}} - \frac{2}{3}(12-\delta_{kl}) \right]\left(|e^2A_{2L}|^2 + |e^2A_{2R}|^2 \right) \nonumber \\
& \qquad  + \frac{4e^2}{3} {\rm Re}\bigg\{ A_{2R}\left[(1+2\delta_{kl})g_4^\ast + g_6^\ast\right] + A_{2L}\left[(1+2\delta_{kl})g_3^\ast + g_5^\ast \right] \bigg\} \bigg].
\label{eq:3-body-decay}
\end{align}
In the case of $\ell^-_i \to \ell^+_k\ell^-_n\ell^-_n$ decay (i.e., $k\neq n$), only box diagrams contribute. The decay width is found to be \cite{Abada:2014kba}
\begin{align}
\Gamma (\ell^-_i \to \ell^+_k\ell^-_n\ell^-_n) = & \frac{m_{\ell_i}^5}{512\pi^3} \left[ \frac{1}{24}\left(|g_1|^2 + |g_2|^2 \right)+ \frac{2}{3} \left(|\tilde g_3|^2+|\tilde g_4|^2\right) +  \frac{1}{3}\left(|\tilde g_5|^2+|\tilde g_6|^2\right)  \right],
\end{align}
where we have defined $\left.\tilde g_i= g_i\right|_{A_{1L,R}=Z_{L,R}=0}$ for $i=3,...,6$. We present current bounds and projected sensitivities of these processes in Table~\ref{tab:202}.

\begin{table}[t!]
\begin{center}
\begin{tabular}{|c|c|c|}\hline
$\ell_i\to \ell_k\ell_m\ell_n$ & Present bound & Future sensitivity \\\hline\hline

$\mu\to eee$&$1.0\times 10^{-12}$~\cite{BELLGARDT19881} &$\sim 10^{-16}$~\cite{Blondel:2013ia}  \\ \hline
$\tau\to eee$&$2.7\times 10^{-8}$~\cite{Hayasaka:2010np} &$\sim 10^{-9}$~\cite{Aushev:2010bq}  \\ \hline
$\tau\to \mu\mu\mu$&$2.1\times 10^{-8}$~\cite{Hayasaka:2010np} &$\sim 10^{-9}$~\cite{Aushev:2010bq}  \\ \hline

$\tau^-\to e^-\mu\mu$&$2.7\times 10^{-8}$~\cite{Hayasaka:2010np} &$\sim 10^{-9}$~\cite{Aushev:2010bq}  \\ \hline
$\tau^-\to \mu^-ee$&$1.8\times 10^{-8}$~\cite{Hayasaka:2010np} &$\sim 10^{-9}$~\cite{Aushev:2010bq}  \\ \hline
$\tau^-\to e^+\mu^-\mu^-$&$1.7\times 10^{-8}$~\cite{Hayasaka:2010np} &$\sim 10^{-9}$~\cite{Aushev:2010bq}  \\ \hline
$\tau^+\to \mu^+e^-e^-$&$1.5\times 10^{-8}$~\cite{Hayasaka:2010np} &$\sim 10^{-9}$~\cite{Aushev:2010bq}  \\ \hline

\end{tabular}
\end{center}
\caption{Current experimental bounds on the $BR(\ell_i\to \ell_k\ell_m\ell_n)$. Future sensitivities are presented on the last column. }
\label{tab:202}
\end{table}

\subsection{Lepton anomalous magnetic-dipole moment}
This quantity arises from the following interaction
\begin{align}
    T = \frac{e}{2m_\mu} F(q^2) \bar u(p_2) \sigma^{\alpha\beta} iq_\beta u(p_1) \epsilon_\alpha(q),
\end{align}
with $q=p_2-p_1$ and $\Delta a_\ell=F(q^2=0)$, evaluated at one-loop penguin diagrams. This gives
\begin{align}
\Delta a_\ell = -\frac{3}{16\pi^2} \sum_{\lambda,\phi} \frac{m_\ell^2}{M_\phi^2} & \bigg\{ \left[|\lambda^{\ell j}_L|^2  + |\lambda^{\ell j}_R|^2 \right]\left[Q_qF_1(x_j)+Q_\phi F_2(x_j) \right] \nonumber \\
& + \frac{m_{q_j}}{m_{\ell}} {\rm Re}(\lambda^{\ell j}_L\lambda^{\ell j\ast}_R) \left[ Q_q F_3(x_j)+ Q_\phi F_4(x_j) \right] \bigg\}.
\label{eq:AMM}
\end{align}
\subsection{$\mu$-$e$ conversion in nuclei}
For this process, we are interested in the so-called coherent processes, that is, no change in nucleon state during the transition happens. The relevant interactions, therefore, can be written as \cite{Kitano:2002mt}
\begin{align}
{\cal L}_{\rm eff} = &~ -m_\mu \bar \mu \sigma^{\alpha\beta} \left(\tfrac{e}{2}A_{2R}P_L + \tfrac{e}{2}A_{2L}P_R \right) e F_{\alpha\beta} \nonumber \\
&~ -\frac{1}{4} \left[ \left( g_{RS}^{(q)} \bar \mu P_R e + g_{LS}^{(q)} \bar \mu P_L e \right) \bar qq  +  \left( g_{RV}^{(q)} \bar \mu \gamma^\alpha P_R e + g_{LV}^{(q)} \bar \mu \gamma^\alpha P_L e \right) \bar q\gamma_\alpha q \right] + {\rm h.c.}
\end{align}
The corresponding Wilson's coefficients are given by
\begin{align}
&g_{RS}^{(q)} = T_{RL}^{(q)}; \quad g_{LS}^{(q)} = T_{LR}, \nonumber \\
&g_{RV}^{(q)} = T_{RR}^{(q)}-2e^2\left[ -2Q_qA_{1R} + Z_R (g^{(q)}_R+g^{(q)}_L) + B_{q} \right], \nonumber \\
&g_{LV}^{(q)} = T_{LL}^{(q)}-2e^2\left[ -2Q_qA_{1L} + Z_L (g^{(q)}_R+g^{(q)}_L) + B_{q} \right], 
\end{align}
with
\begin{align}
&T_{XY}^{(q)} = \sum_{\lambda,\phi}\frac{\lambda_X^{2q}\lambda_Y^{1q\ast}}{M_\phi^2}, \\
&B_{q} = \frac{1}{16\pi^2} \sum_{\lambda,\phi} \frac{\lambda_L^{2j}\lambda_L^{1j\ast}}{M_\phi^2} \left [(\lambda_L^\dagger \lambda_L)^{qq} + (\lambda_R^\dagger\lambda_R)^{qq} \right] b_1(x_j,0).
\end{align}

Using these expressions, we can calculate the $\mu$-$e$ transition rate
\begin{eqnarray}
\Gamma_{\mu-e~\rm conv.} &=& \tfrac{1}{4} \left| \tfrac{e}{2}A_{2R}D + \tilde g_{LS}^{(p)}S^{(p)} + \tilde g_{LS}^{(n)}S^{(n)} + \tilde g_{LV}^{(p)}V^{(p)} + \tilde g_{LV}^{(n)} V^{(n)} \right|^2 \nonumber \\
&& + (L\leftrightarrow R),
\end{eqnarray}
where $D,S^{(p,n)},V^{(p,n)}$ are the overlap integrals for each operator. Their values in the unit of $m_\mu^{5/2}$ are given in \cite{Kitano:2002mt}. The effective couplings $\tilde g^{(N)}_{LK,RK}$ with $K=S,V$ are given by
\begin{align}
\tilde g_{LK,RK}^{(N)} = \sum_q G_K^{(q,N)} g_{LK,RK}^{(q)}.
\end{align}
The sum runs over all quark flavors for $K=S$ and runs over valence quarks for $K=V$. Numerically, $G_V^{(u,p)}=G_V^{(d,n)}=2$, $G_V^{(d,p)}=G_V^{(u,n)}=1$, and $G_S^{(u,p)}=G_S^{(d,n)}=5.1$, $G_S^{(d,p)}=G_S^{(u,n)}=4.3$, $G_S^{(s,p)}=G_S^{(s,n)}=2.5$, while those of heavy quarks are negligible. As in the previous LFV cases, we present current bounds and future sensitivities of this process in Table \ref{tab:203}.

\begin{table}[t!]
\begin{center}
\begin{tabular}{|c|c|c|}\hline
Nucleus & Present bound & Future sensitivity \\\hline\hline

Gold &$7\times 10^{-13}$~\cite{SINDRUMII:2006dvw} &$-$  \\ \hline
Titanium &$4.3\times 10^{-12}$~\cite{DOHMEN1993631}&$\sim 10^{-18}$~\cite{unPUB}  \\ \hline
Aluminum &$-$&$10^{-15}-10^{-18}$~\cite{Pezzullo:2017iqq}  \\ \hline

\end{tabular}
\end{center}
\caption{Current experimental bounds on the $BR(\mu - e)$ conv. in the nuclei. Future sensitivities are presented on the last column. }
\label{tab:203}
\end{table}

\section{Results}\label{sec:results}
In the previous section, we derived all charged lepton flavor violating processes that can be used to constrain the model's parameter space. Since addressing flavor anomalies demands TeV-scale leptoquarks with some of the couplings being of order one, bounds from lepton flavor violating processes provide the most stringent constraints on the Yukawa couplings, which we explore in this section in great details.

\subsection{Case studies}
The neutrino mass formula given in Eq.~\eqref{nu-mass-up} (or Eq.~\eqref{nu-mass-down}) consists of four different Yukawa couplings $y^{L,R}$ and $f^{L,R}$, which are a priori arbitrary $3\times 3$ matrices. The parameter space is quite broad; therefore, we choose a few specific benchmark scenarios and perform a detailed numerical analysis. Since the terms in the second and the third lines in the neutrino mass formula Eq.~\eqref{nu-mass-up} (or Eq.~\eqref{nu-mass-down}) are proportional to $m_\tau/m_t$, for Yukawa couplings of a similar order, these terms can be completely neglected. This is why, for our numerical study, we stick to the simplified scenario where $f^R$ and $y^R$ also provide sub-leading contributions to LFV unless otherwise explicitly mentioned. To further reduce the parameters, we assume vanishing Yukawa couplings with the first generation quarks, i.e., $y^L_{1i}, f^L_{1i}=0$.

Among the few predictive cases that we consider, in the following, we first discuss the most minimal scenario consisting of six non-zero Yukawa parameters $y^L_{3j}, f^L_{3j}\neq 0$ that provides an excellent fit to the neutrino oscillation data. As will be discussed later in the text, further parameter space reduction fails to fit neutrino observables with their respective $2\sigma$ values. Considering the loop integral behavior discussed above and the suppression of $m_\tau/m_t$, in the case of no large hierarchy among Yukawa couplings, it is an excellent approximation to keep only the third generation of quarks. Then the neutrino mass matrix formula, in this case, becomes
\begin{align}
    (\mathcal{M}_\nu)_{ji} \simeq \frac{3g^2m_t}{\sqrt{2}(16\pi^2)^2}
 \left[ y^L_{3j} f^L_{3i} + f^L_{3j} y^L_{3i} \right] \hat I_{j33}.
\label{nu-mass}
\end{align}
The above formula applies to both the up-quark and down-quark mass diagonal bases. This is because, in the limit we are working, in the up-quark mass diagonal basis Eq.~\eqref{nu-mass-up}, the loop integrals are flavor independent. On the other hand, in the down-quark mass diagonal basis Eq.~\eqref{nu-mass-down}, the remaining factor is $V_{tb}\approx 1$. One can see from this formula that the neutrino mass matrix is reduced into a rank two matrix, whose determinant vanishes. Thus, this specific texture predicts that one of the neutrinos is massless, although both neutrino mass orderings, i.e., normal hierarchy (NH) and inverted hierarchy (IH), can be admitted. The neutrino mass matrix in this form can nicely fit oscillation data.

Before diving into numerics, we first demonstrate that the undetermined Yukawa couplings appearing in the above neutrino mass formula can be fully expressed in terms of neutrino observables and as a function of LQ masses and mixing parameters.  By following  the parametrization described in  \cite{Cordero-Carrion:2018xre,Cordero-Carrion:2019qtu} (for alternative parameterizations, see also, Ref. \cite{Cai:2014kra,Hagedorn:2018spx}), we determine these Yukawa couplings appearing in Eq.~\eqref{nu-mass}. To do so, the neutrino mass matrix is  diagonalized as follows:
\begin{align}
&\mathcal{M}_\nu= U^* \begin{pmatrix} m_1&0&\\0&m_2&0\\0&0&m_3 \end{pmatrix}   U^\dagger,
\end{align}
where $U$ is the Pontecorvo-Maki-Nakagawa-Sakata (PMNS) mixing matrix and $m_i$ are neutrino mass eigenvalues given by
\begin{align}
&m_1=0,\;m_2=\sqrt{\Delta m^2_{21}},\;m_3=\sqrt{\Delta m^2_{31}},
\end{align}
for NH and
\begin{align}
&m_1= \sqrt{-\Delta m^2_{32}-\Delta m^2_{21}},\; m_2=\sqrt{-\Delta m^2_{32}},\; m_3=0,
\end{align}
for IH. Now the neutrino mass matrix given in Eq. \eqref{nu-mass} can be re-written as
\begin{align}
&\mathcal{M}_\nu= a_0 \left( Y^{T}_a \hat{m} Y_b+Y^{T}_b\hat{m} Y_a \right), 
\\&
a_0=\frac{3g^2}{\sqrt{2}(16\pi^2)^2},\;\; 
\hat{m}= m_t\hat I_{j33}.
\end{align}
Here, $Y_a=y^L_{3i}$ and $Y_b=f^L_{3i}$ are row matrices.
Utilizing this form, the two unknown Yukawa coupling matrices can be entirely determined by the known values \cite{Esteban:2020cvm} of neutrino observables, SM fermion masses, and as a function of scalar masses and mixings that run through the loops. For NH, we have
\begin{align}
&Y^{T}_a=\frac{16\pi^2}{3^{1/2}2^{1/4}g} \begin{pmatrix}
i\;r_2\;U^*_{12}+r_3\;U^*_{13}\\
i\;r_2\;U^*_{22}+r_3\;U^*_{23}\\
i\;r_2\;U^*_{32}+r_3\;U^*_{33}
\end{pmatrix},\\
&Y^{T}_b=\frac{16\pi^2}{3^{1/2}2^{1/4}g} \begin{pmatrix}
-i\;r_2\;U^*_{12}+r_3\;U^*_{13}\\
-i\;r_2\;U^*_{22}+r_3\;U^*_{23}\\
-i\;r_2\;U^*_{32}+r_3\;U^*_{33}
\end{pmatrix},
\end{align}
Similarly, for IH, the solution for $Y^{a,b}$ takes the following forms
\begin{align}
&Y^{T}_a=\frac{16\pi^2}{3^{1/2}2^{1/4}g} \begin{pmatrix}
r_1\;U^*_{11}+i\;r_2\;U^*_{12}\\
r_1\;U^*_{21}+i\;r_2\;U^*_{22}\\
r_1\;U^*_{31}+i\;r_2\;U^*_{32}
\end{pmatrix},\\
&Y^{T}_b=\frac{16\pi^2}{3^{1/2}2^{1/4}g} \begin{pmatrix}
r_1\;U^*_{11}-i\;r_2\;U^*_{12}\\
r_1\;U^*_{21}-i\;r_2\;U^*_{22}\\
r_1\;U^*_{31}-i\;r_2\;U^*_{32}
\end{pmatrix}.
\end{align}
In both hierarchies,  we define $r_i=\left( m_i/\hat{m} \right)^{1/2}$.

This parametrization is sometimes useful to fix the undetermined Yukawa parameters of the theory. In our detailed numerical analysis, we consider not only the benchmark (BM) scenario as mentioned above but also a few variations of it that include reducing as well as extending the number of parameters. Particularly, all case studies we examine are summarized in the following:
\begin{enumerate}
\item[$\bullet$] Texture given in Eq.~\eqref{nu-mass} with $f^L_{3j}, y^L_{3j}\neq 0$. We study both NH and IH, which we label as \textbf{NH-I} and \textbf{IH-I}, respectively. Both cases provide a good fit to neutrino data.   

\item[$\bullet$] A more minimal variation of the scenario mentioned above is to choose at least one of $f^L_{3j}=0$ or  $y^L_{3j}= 0$, leading to vanishing $({\cal M}_{\nu})_{jj}$.  Coupled with the fact that the lightest neutrino is massless, this restriction clearly does not work for NH. Interestingly, for IH, the case with  $f^L_{32}=0$ or  $y^L_{32}= 0$ can still be fitted within $3\sigma$ experimental values of the neutrino observables. We demonstrate this by choosing $f^L_{32}=0$ and label it as \textbf{IH-II}.   

\item[$\bullet$] In the cases mentioned above, with all zero entries in the first and the second rows, the ($31$)-entries of both coupling matrices are required to be comparable with the other entries to provide a good fit, which subsequently leads to large $\mu\to e \gamma$. Consequently, these cases demand large LQ masses to be consistent with the non-observation of LVF. In search for a minimal texture that is also compatible with $\sim \mathcal{O}(1)$ TeV LQs, we explore a scenario with $f^L_{31}=0$ but introduce nonzero couplings $f^L_{2j}, y^L_{2j}$ for at least one $j$. Now, although $({\cal M}_\nu)_{11}=0$, the determinant of the neutrino mass matrix is no longer zero, so a viable neutrino fit, which is compatible with NH, can be obtained. (A vanishing (11)-element of neutrino mass matrix cannot be realized in IH case.) For demonstration purpose, we choose to have nonzero $f^L_{23}$ and $y^L_{23}$, while other  $y^L_{2j},f^L_{2j}$ are simply set to zero. The two working benchmarks are labeled  as \textbf{NH-II} (up-diagonal basis) and \textbf{NH-III} (down-diagonal basis). 
\end{enumerate}

\subsection{Numerical analysis}
Our numerical study is based on $\chi^2$ analysis, and the $\chi^2$-function is defined as
\begin{align}
\chi^2= \sum_i  \left( \frac{T_{i}-E_{i}}{\sigma_i}\right)^2,
\end{align}
where $\sigma_i$ represents experimental $1\sigma$ uncertainty; $T_{i}$ and  $E_{i}$ represent the theoretical prediction and the experimental central value for the $i$-th observable, respectively. In the above equation, $i$ is summed over five observables: two neutrino mass squared differences and three mixing angles.  For the simplicity of our work, we consider all parameters to be real; hence, we do not attempt to fit the CP-violating Dirac phase in the neutrino sector, which can be trivially done by turning on phases of these couplings. Neutrino oscillation data used in our fit are summarized in Table \ref{tab:nuEXP}.  Once a good fit to data is obtained from $\chi^2$ analysis, we perform a Markov chain Monte Carlo (MCMC) analysis to explore the parameter space (consistent with neutrino observables) and inspect lepton flavor violation for which we varied the non-zero Yukawa couplings and LQ masses in the ranges $[-1,1]$ and $[1,100]$ TeV, respectively. 

\begin{table}[t!] 
\centering
\begin{footnotesize}
    \begin{tabular}{c|l|cc|cc}
      \hline\hline
      \multirow{11}{*}{} &
      & \multicolumn{2}{c|}{Normal Ordering}
      & \multicolumn{2}{c}{Inverted Ordering}
      \\
      \cline{3-6}
      && bfv $\pm 1\sigma$ & $3\sigma$ range
      & bfv $\pm 1\sigma$ & $3\sigma$ range
      \\
      \cline{2-6}
      \rule{0pt}{4mm}\ignorespaces
      & $\sin^2\theta_{12}$
      & $0.304_{-0.012}^{+0.013}$ & $0.269 \to 0.343$
      & $0.304_{-0.012}^{+0.012}$ & $0.269 \to 0.343$
      \\[3mm]
      & $\sin^2\theta_{23}$
      & $0.573_{-0.023}^{+0.018}$ & $0.405 \to 0.620$
      & $0.578_{-0.021}^{+0.017}$ & $0.410 \to 0.623$
      \\[3mm]
      & $\sin^2\theta_{13}$
      & $0.02220_{-0.00062}^{+0.00068}$ & $0.02034 \to 0.02430$
      & $0.02238_{-0.00062}^{+0.00064}$ & $0.02053 \to 0.02434$
      \\[3mm]
      & $\frac{\Delta m^2_{21}}{10^{-5}}$ eV$^2$
      & $7.42_{-0.20}^{+0.21}$ & $6.82 \to 8.04$
      & $7.42_{-0.20}^{+0.21}$ & $6.82 \to 8.04$
      \\[3mm]
      & $\frac{\Delta m^2_{3\ell}}{10^{-3}}$ eV$^2$
      & $2.515_{-0.028}^{+0.028}$ & $2.431 \to 2.599$
      & $-2.498_{-0.029}^{+0.028}$ & $-2.584 \to -2.413$
      \\[2mm]
      \hline\hline
      \end{tabular}
\end{footnotesize}
\caption{Neutrino oscillation parameters taken from Ref.~\cite{Esteban:2020cvm}. Here, $\Delta m^2_{31} > 0$ for NH and
    $\Delta m^2_{32} < 0$ for IH. Here 'bfv' represents best fit values obtained from global fit~\cite{Esteban:2020cvm}. }\label{tab:nuEXP}
\end{table}

Sample fits obtained from our numerical procedure that is consistent with neutrino observables are presented below for each of the cases listed above (here, we have defined $m_0=a_0 \hat m$):
\begin{align}
&\textrm{\bf IH-I:}\;\; m_0= 0.0576\; \textrm{eV},
\\
&f^L=\left(
\begin{array}{ccc}
 0 & 0 & 0 \\
 0 & 0 & 0 \\
 -0.1962 & -0.6212 & -0.7730 \\
\end{array}
\right),\;\;\;
y^L=\left(
\begin{array}{ccc}
 0 & 0 & 0 \\
 0 & 0 & 0 \\
 0.8239 & -0.2074 & -0.0508 \\
\end{array}
\right). \label{IH-I}
\\
&\textrm{\bf IH-II:}\;\; m_0= 0.0592\; \textrm{eV},
\\
&f^L=\left(
\begin{array}{ccc}
 0 & 0 & 0 \\
 0 & 0 & 0 \\
 -0.7134 & 0 & -0.1701 \\
\end{array}
\right),\;\;\;
y^L=\left(
\begin{array}{ccc}
 0 & 0 & 0 \\
 0 & 0 & 0 \\
 0.1931 & 0.7401 & -0.8472 \\
\end{array}
\right). \label{IH-II}
\\
&\textrm{\bf NH-I:}\;\; m_0= 0.0418\; \textrm{eV},
\\
&f^L=\left(
\begin{array}{ccc}
 0 & 0 & 0 \\
 0 & 0 & 0 \\
 -0.0680 & -0.4836 & -0.8209 \\
\end{array}
\right),\;\;\;
y^L=\left(
\begin{array}{ccc}
 0 & 0 & 0 \\
 0 & 0 & 0 \\
 0.2549 & -0.6442 & -0.2462 \\
\end{array}
\right). \label{NH-I}
\\
&\textrm{\bf NH-II:}\;\; m_0= 24.0268\; \textrm{eV},
\\
&f^L=\left(
\begin{array}{ccc}
 0 & 0 & 0 \\
 0 & 0 & -0.9747 \\
 0 & 0.00974 & -0.03489 \\
\end{array}
\right),\;\;\;
y^L=\left(
\begin{array}{ccc}
 0 & 0 & 0 \\
 0 & 0 & 0.9871 \\
 0.01048 & -0.05672 & -0.08842 \\
\end{array}
\right). \label{NH-II}
\\
&\textrm{\bf NH-III:}\;\; m_0= 8.4371\; \textrm{eV},
\\
&f^L=\left(
\begin{array}{ccc}
 0 & 0 & 0 \\
 0 & 0 & -0.4931 \\
 0 & 0.0199 & -0.01547 \\
\end{array}
\right),\;\;\;
y^L=\left(
\begin{array}{ccc}
 0 & 0 & 0 \\
 0 & 0 & 0.4419 \\
 0.05253 & -0.05396 & -0.1469 \\
\end{array}
\right). \label{NH-III}
\end{align}
The corresponding fit values of neutrino observables are collected in Table \ref{tab:NUfit}, and the resulting neutrino mass matrices are shown in the following: 
\begin{align}
&\mathcal{M}^\textrm{IH-I}_\nu=
\left(
\begin{array}{ccc}
 -0.01864 & -0.02716 & -0.03614 \\
 -0.02716 & 0.01485 & 0.01106 \\
 -0.03614 & 0.01106 & 0.004534 \\
\end{array}
\right)\; \textrm{eV}, 
\\
&\mathcal{M}^\textrm{IH-II}_\nu=
\left(
\begin{array}{ccc}
 -0.01633 & -0.03129 & 0.03387 \\
 -0.03129 & 0 & -0.007460 \\
 0.03387 & -0.007460 & 0.01708 \\
\end{array}
\right)\; \textrm{eV}, 
\\
&\mathcal{M}^\textrm{NH-I}_\nu=
\left(
\begin{array}{ccc}
 -0.001451 & -0.003323 & -0.008053 \\
 -0.003323 & 0.02606 & 0.02710 \\
 -0.008053 & 0.02710 & 0.01691 \\
\end{array}
\right)\; \textrm{eV}, 
\\
&\mathcal{M}^\textrm{NH-II}_\nu=
\left(
\begin{array}{ccc}
 0 & 0.002452 & -0.008786 \\
 0.002452 & -0.02655 & 0.02686 \\
 -0.008786 & 0.02686 & -0.01697 \\
\end{array}
\right)\; \textrm{eV}, 
\\
&\mathcal{M}^\textrm{NH-III}_\nu=
\left(
\begin{array}{ccc}
 0 & 0.00885 & 0.002236 \\
 0.00885 & -0.01818 & -0.02705 \\
 0.002236 & -0.02705 & -0.02534 \\
\end{array}
\right)\; \textrm{eV}. 
\end{align}

\begin{table}[t!]
\centering
{\footnotesize
\resizebox{0.9\textwidth}{!}{
\begin{tabular}{|c|c|c|c|c|c|}
\hline
{\bf Quantity} & {\bf IH-I}  & {\bf IH-II}  & {\bf NH-I} & {\bf NH-II} & {\bf NH-III} \\[3pt]
\hline\hline
$\sin^2\theta_{12}$
&0.304&0.327&0.305&0.304&0.304 \\[3pt]\hline

$\sin^2\theta_{23}$
&0.578&0.593&0.572&0.574&0.448 \\[3pt]\hline

$\sin^2\theta_{13}$
&0.02239&0.022611&0.02223&0.02237&0.02234 \\[3pt]\hline

$\Delta m^2_{21}\times 10^{5}$ eV$^2$
&7.425&7.408&7.425&7.4111&7.428 \\[3pt]\hline

$\Delta m^2_{3l}\times 10^{3}$ eV$^2$
&-2.498& -2.498&2.515&2.514&2.513 \\[3pt]\hline

\end{tabular}
}
\caption{Neutrino mass-squared differences and mixing angles obtained from fits for all the cases studied in this work. Here we have defined $\Delta m^2_{3l}=\Delta m^2_{31}>0$ for NH and $\Delta m^2_{3l}=\Delta m^2_{32}<0$ for IH.   }
    \label{tab:NUfit}
    }
\end{table}

\begin{figure}[t!]
	\centering
\includegraphics[width=6cm]{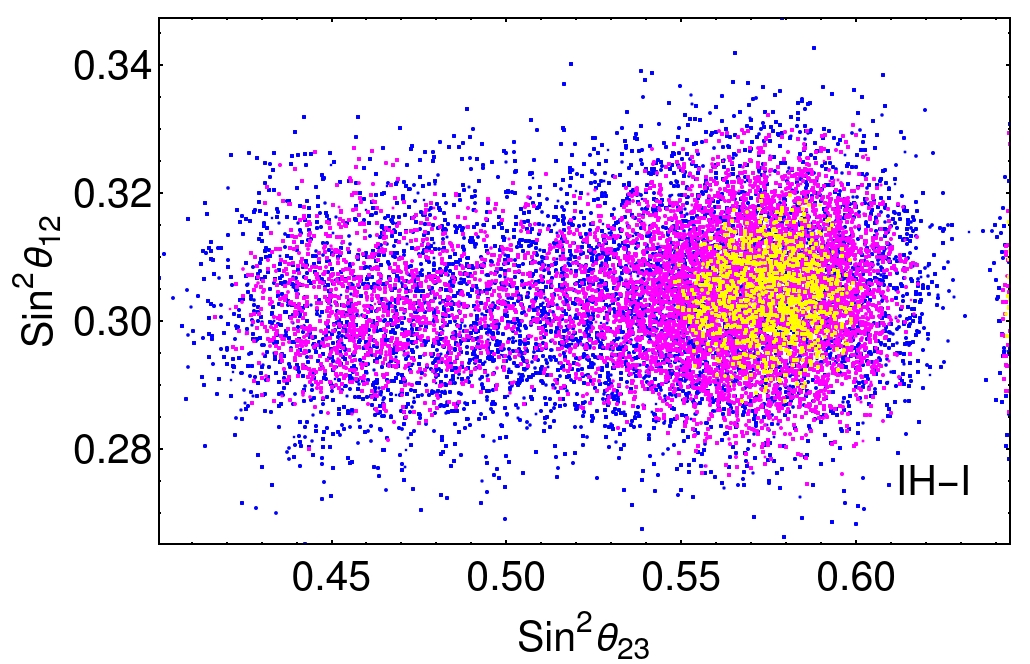}
\includegraphics[width=6cm]{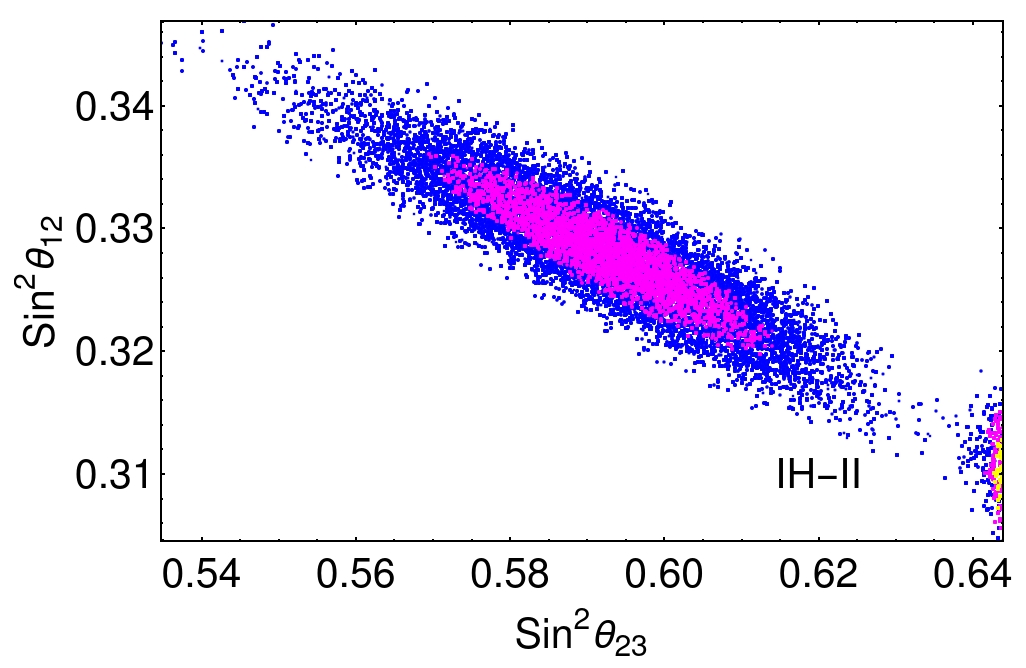}
\includegraphics[width=6cm]{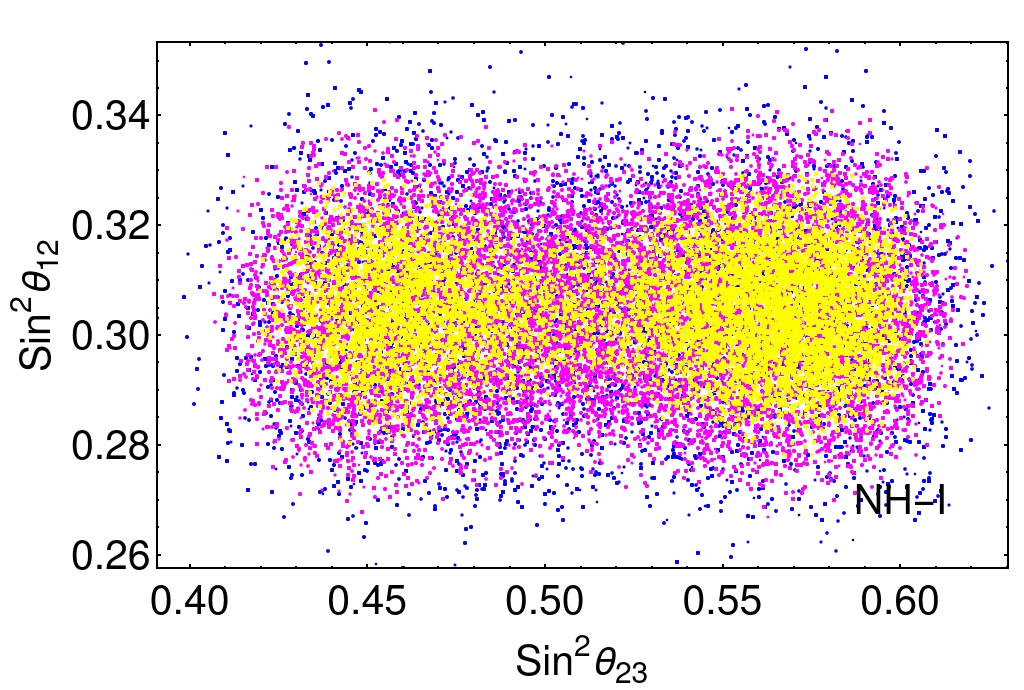}
\includegraphics[width=6cm]{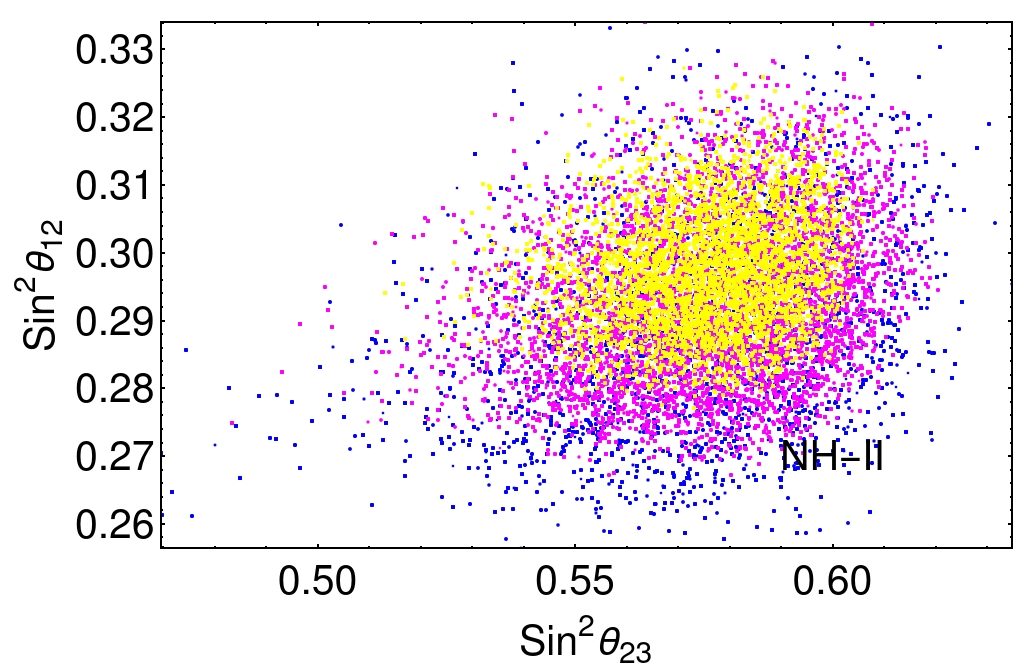}
\includegraphics[width=7cm]{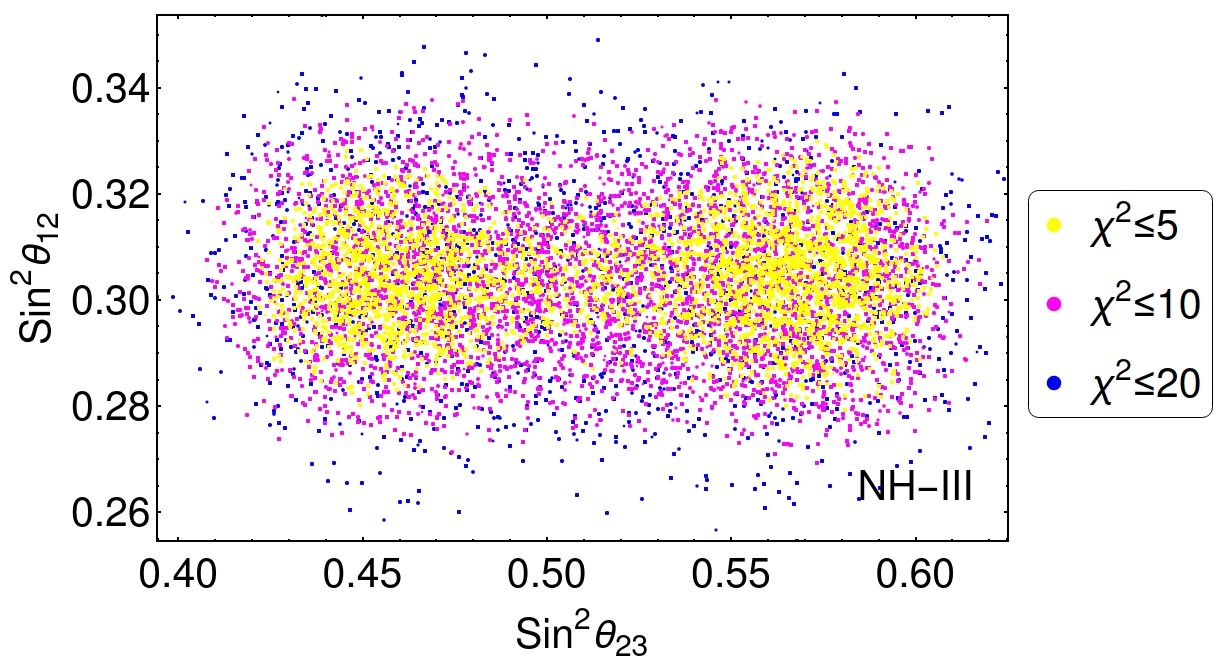}
	\caption{Correlations between $\sin^2\theta_{12}$ and $\sin^2\theta_{23}$ for the five different textures we study. See text for details.}\label{fig:109}
\end{figure}

\begin{figure}[th!]
	\centering
\includegraphics[width=7.5cm]{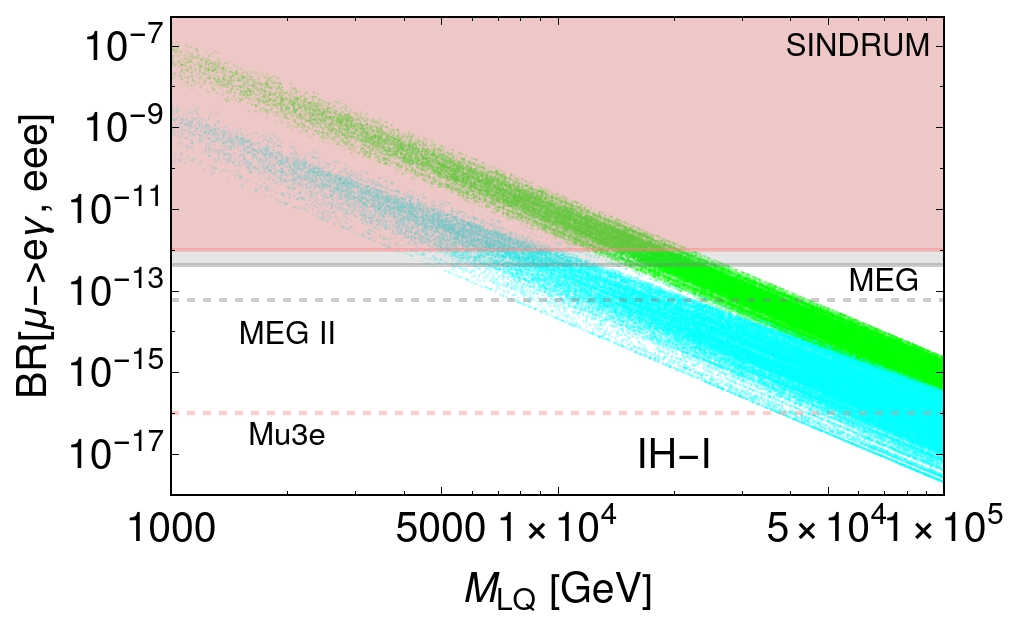}
\includegraphics[width=7.5cm]{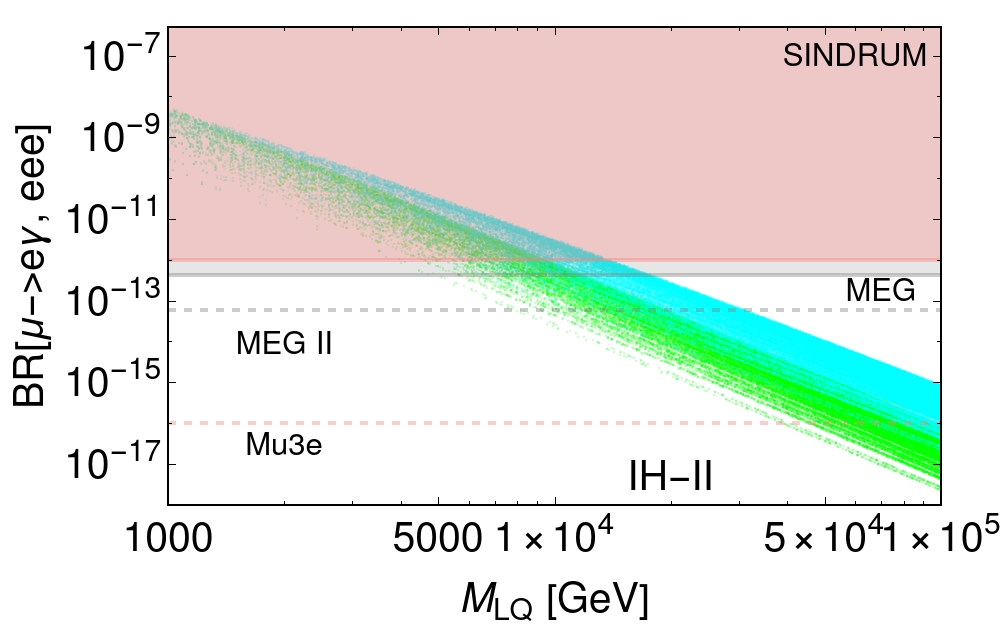}
\includegraphics[width=9.8cm]{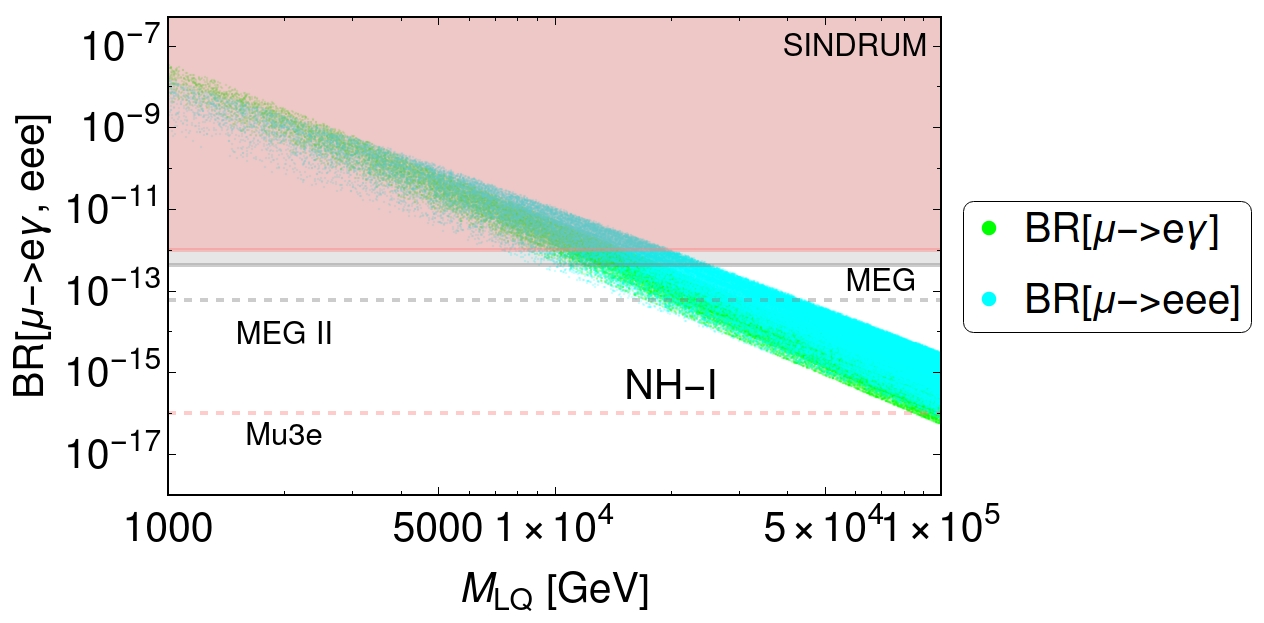}
	\caption{$BR(\mu\to e\gamma)$ and $BR(\mu\to eee)$ as a function of the common LQ mass. Shaded colored regions are ruled out by current data and dotted lines represent future sensitivities.}\label{fig:105}
\end{figure}

\begin{figure}[th!]
	\centering
\includegraphics[width=7.5cm]{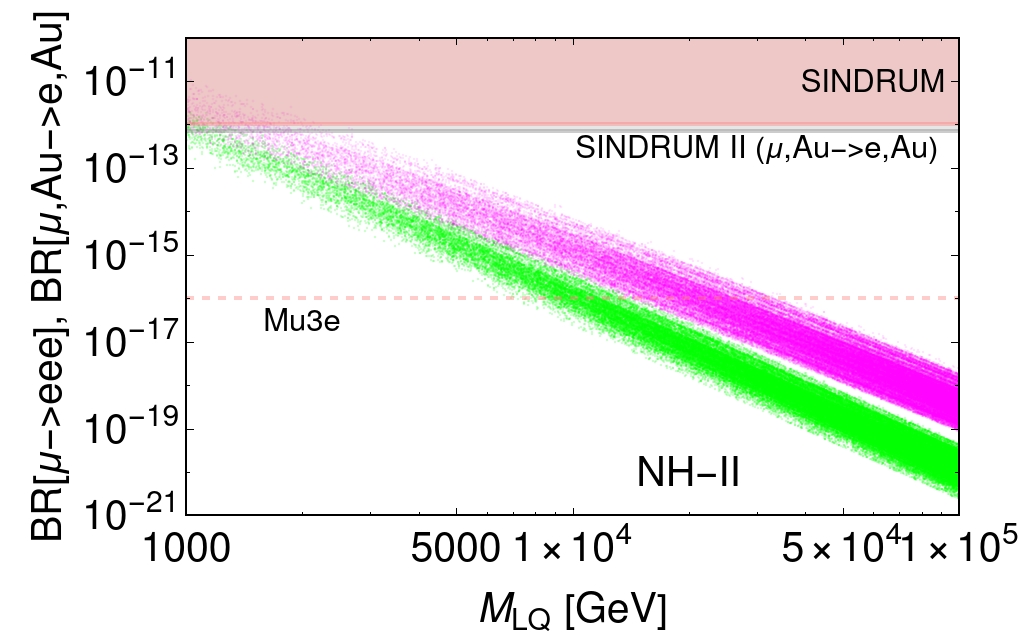}
\includegraphics[width=7.5cm]{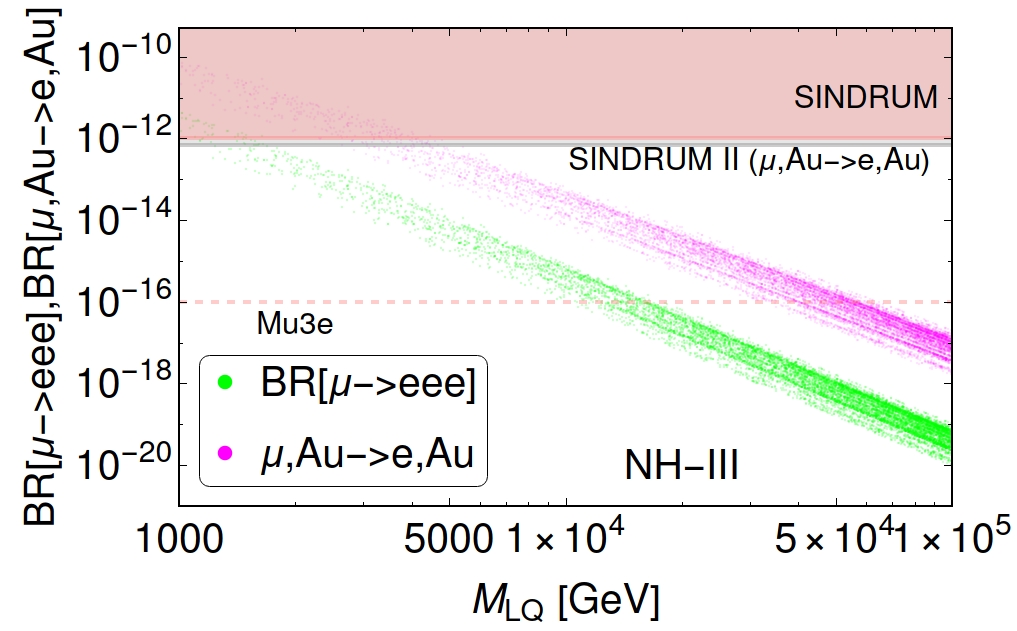}
	\caption{$BR(\mu\to eee)$ and $BR(\mu- e)$ conversion as a function of the common LQ mass. Shaded colored regions are ruled out by current data and dotted lines represent future sensitivities.}\label{fig:106}
\end{figure}

\begin{figure}[t!]
	\centering
\includegraphics[width=7.5cm]{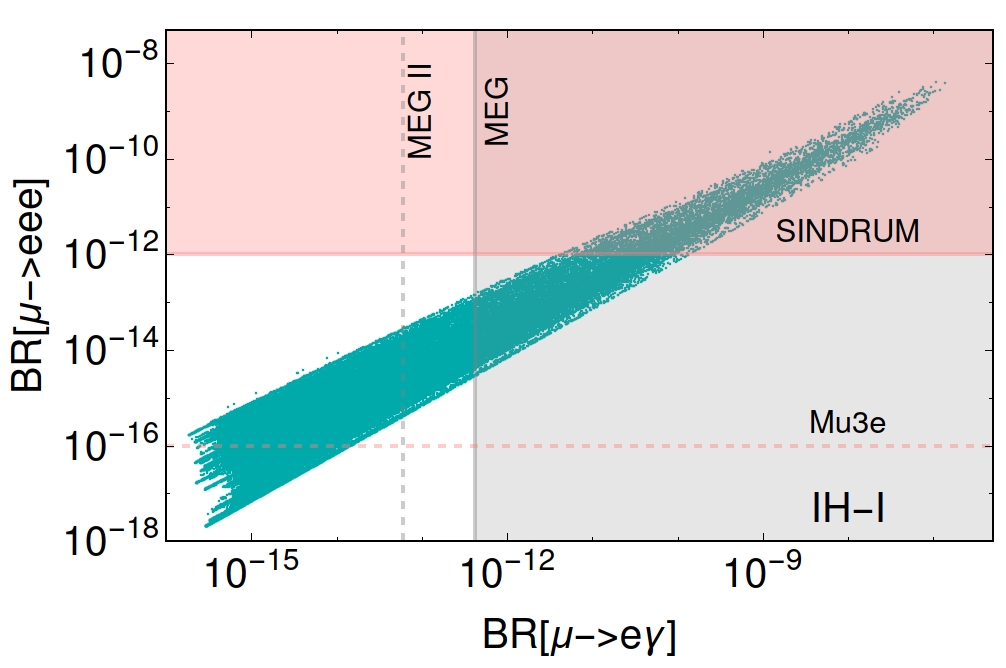}
\includegraphics[width=7.5cm]{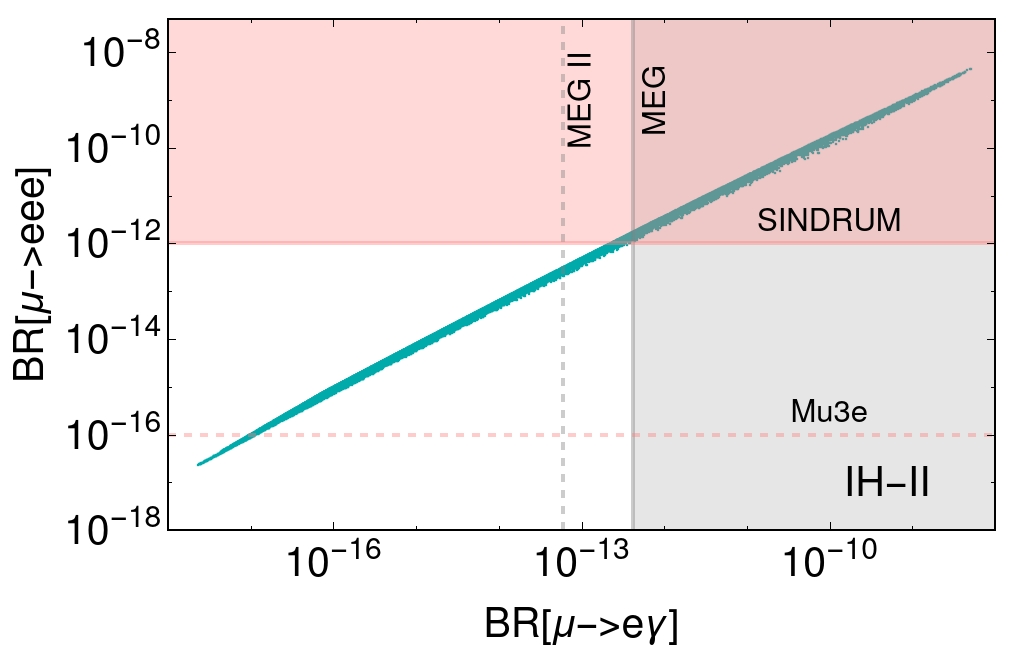}
\includegraphics[width=7.5cm]{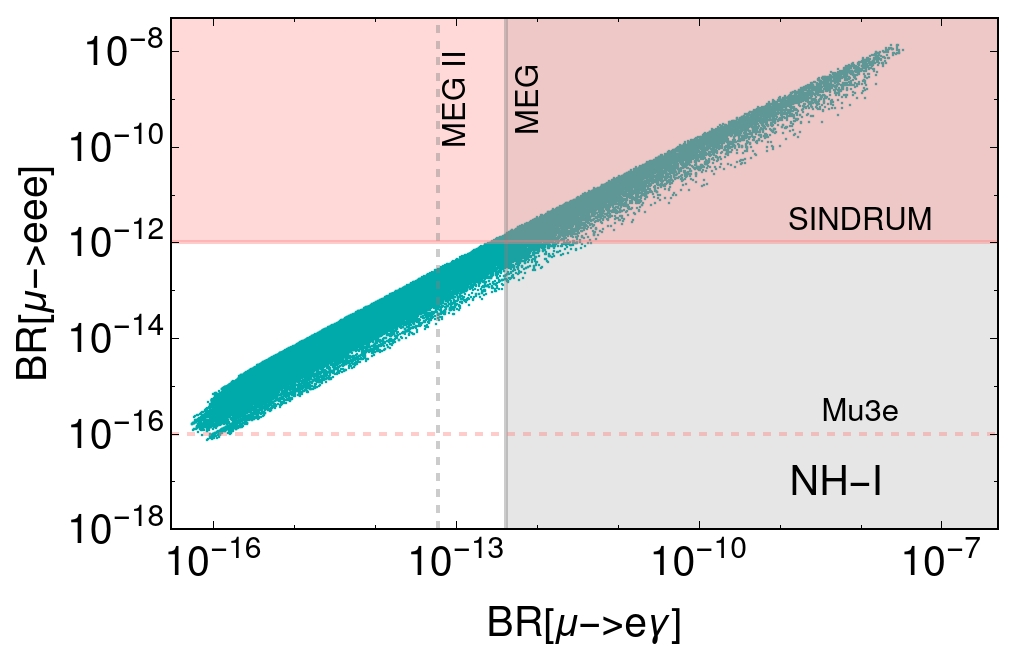}
	\caption{Correlations between $\mu\to e\gamma$ and  $\mu\to eee$. Shaded colored regions are ruled out by current data and dotted lines represent future sensitivities.}\label{fig:102}
\end{figure}

\begin{figure}[t!]
	\centering
\includegraphics[width=7.5cm]{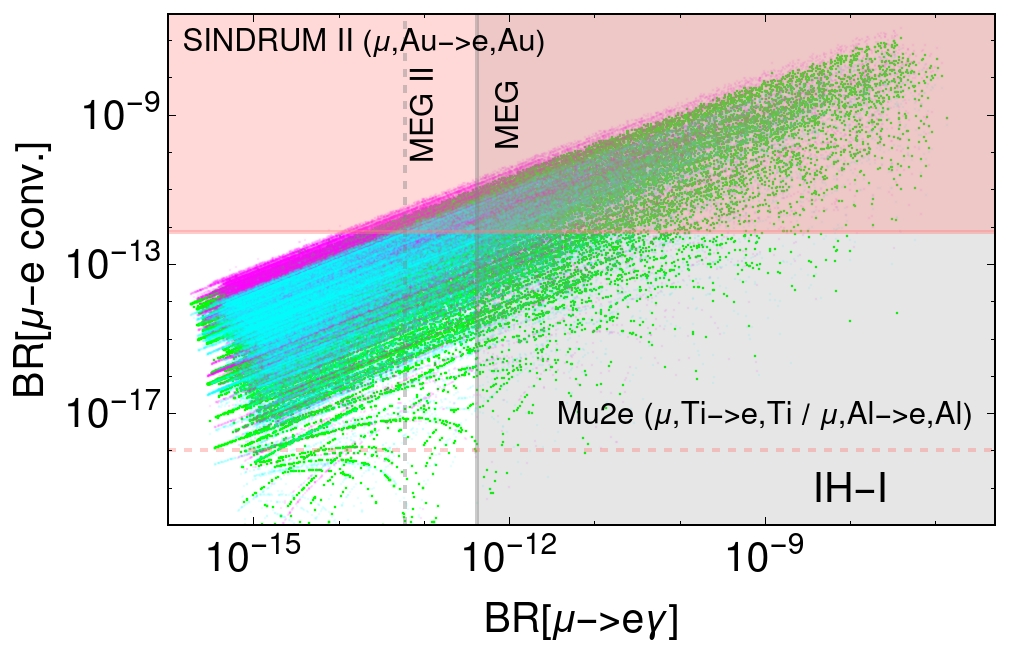}
\includegraphics[width=7.5cm]{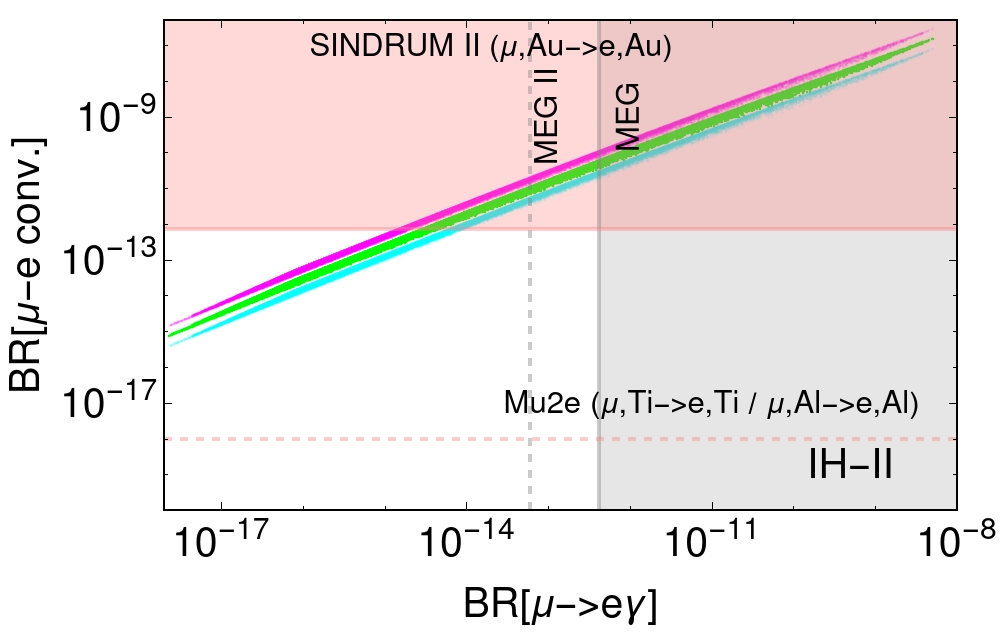}
\includegraphics[width=9.8cm]{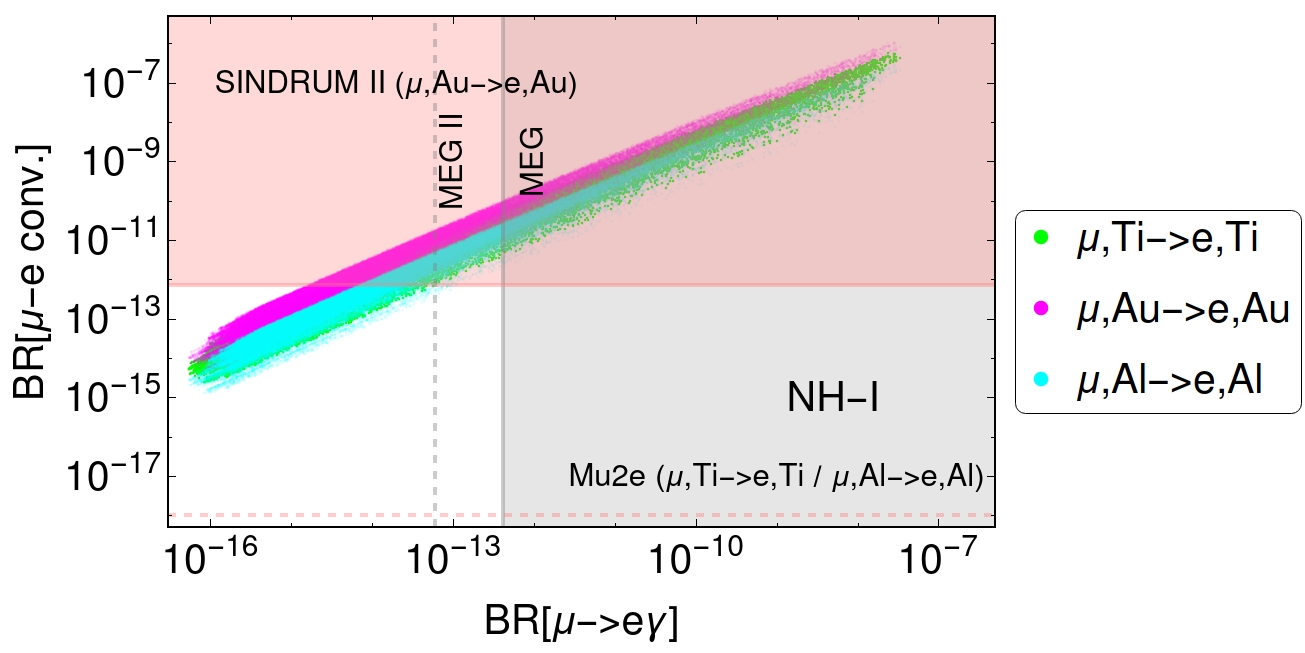}
	\caption{Correlations between $\mu\to e\gamma$ and $\mu- e$ conversion. Shaded colored regions are ruled out by current data and dotted lines represent future sensitivities.}\label{fig:103}
\end{figure}

\begin{figure}[t!]
	\centering
\includegraphics[width=7.5cm]{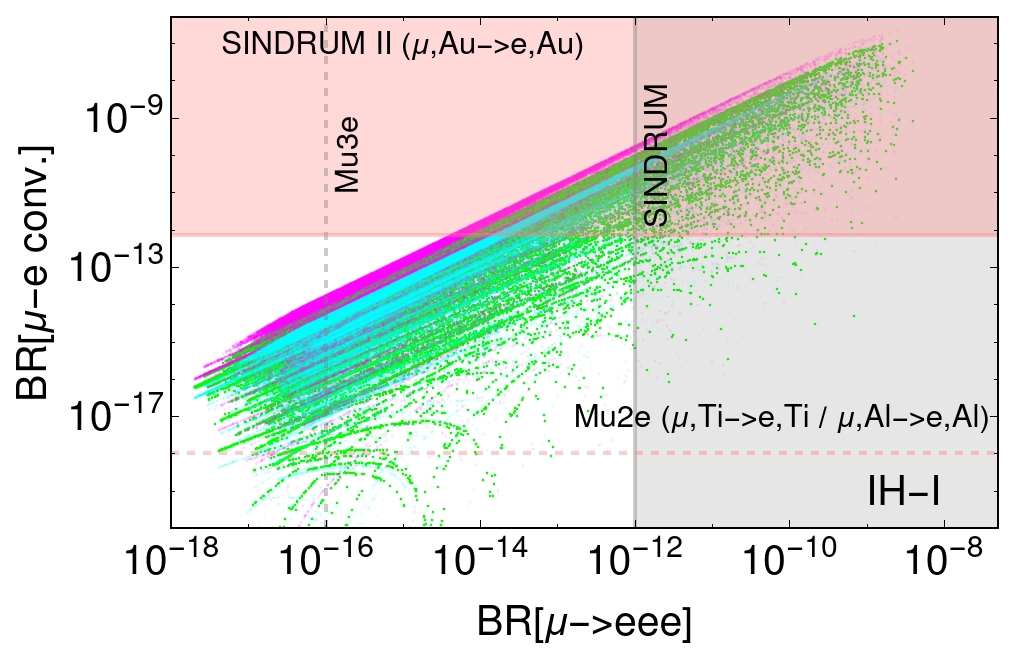}
\includegraphics[width=7.5cm]{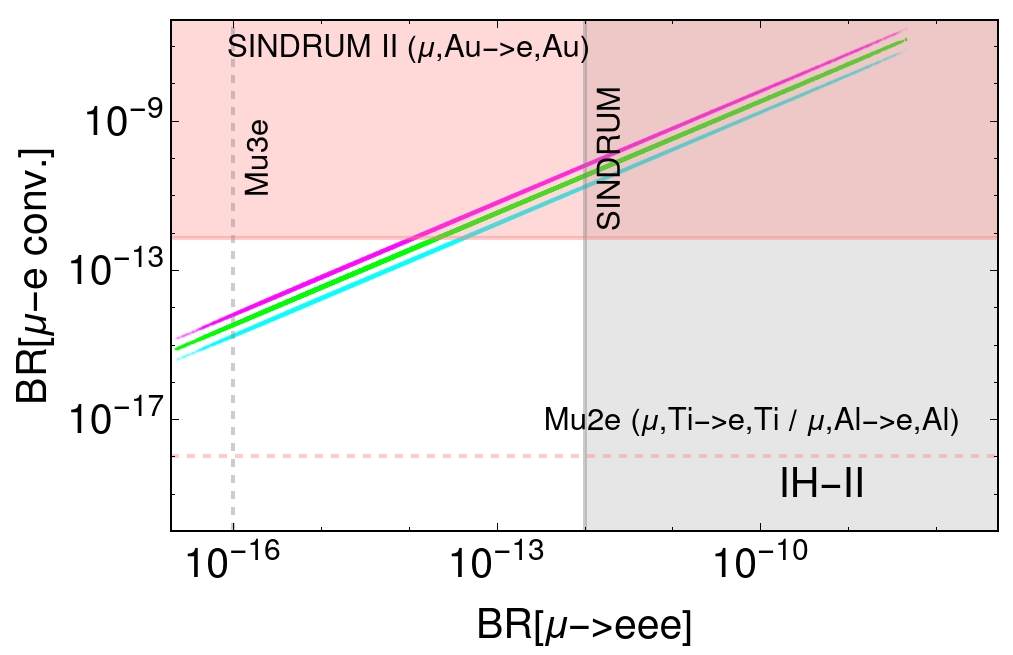}
\includegraphics[width=7.5cm]{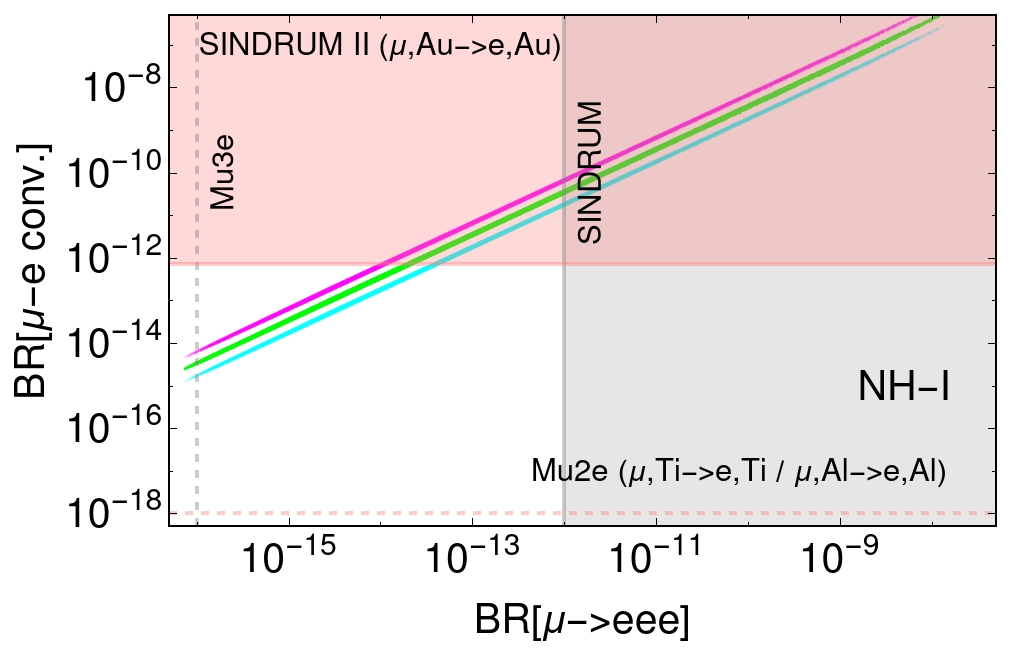}
\includegraphics[width=7.5cm]{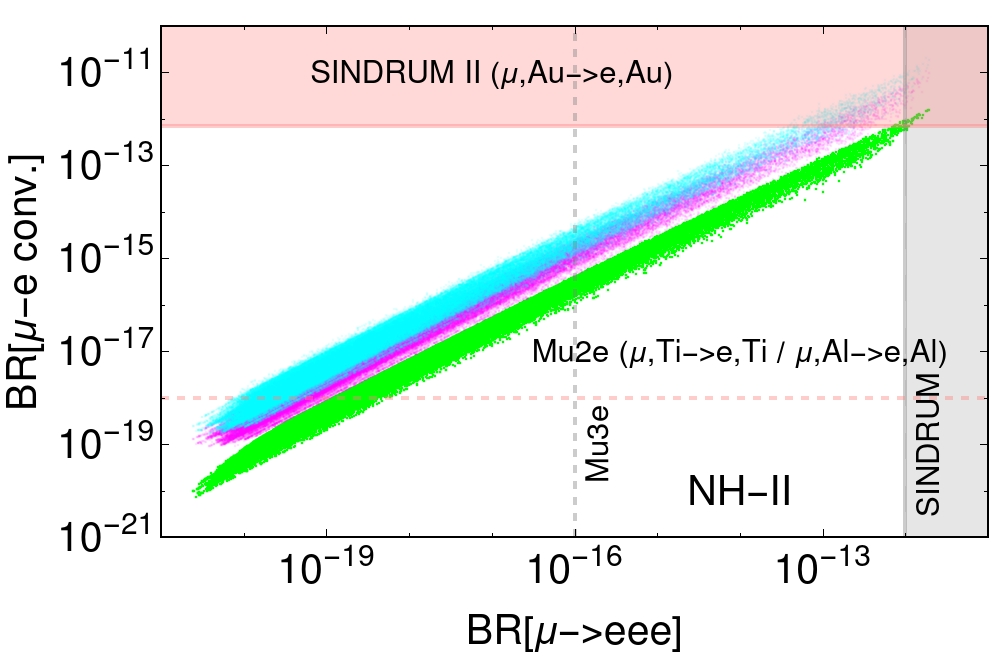}
\includegraphics[width=9.8cm]{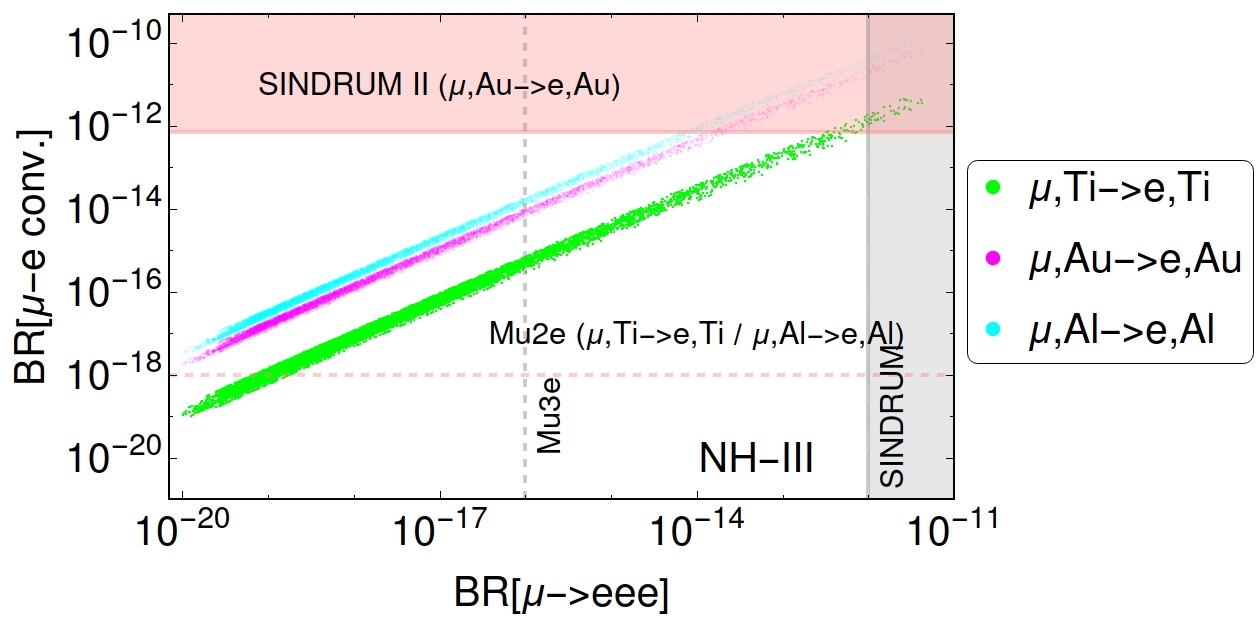}
	\caption{Correlations between $\mu- e$ and $\mu\to eee$. Shaded colored regions are ruled out by current data and dotted lines represent future sensitivities.}\label{fig:104}
\end{figure}

From our detailed numerical scan over the parameter space using MCMC analysis, we obtain interrelationships among various observables: correlations between neutrino mixing parameters, bounds on LQ masses from LFV processes, and correlations among different LVF processes are presented in Figs. \ref{fig:109}-\ref{fig:104}. 

For both the NH and IH cases, a global fit to neutrino oscillation data has two local minima for the mixing angle $\theta_{23}$, the one with $\theta_{23}>45^\circ$ being the lower one~\cite{Esteban:2020cvm} (for those without SuperKamiokande data). Whereas for NH, these two minima are almost identical (in the sense of $\Delta \chi^2$ measure), however, they significantly differ for IH, and $\theta_{23}>45^\circ$ case is highly preferred to $\theta_{23}<45^\circ$. This feature is clearly visible in the upper left panel in Fig.~\ref{fig:109} for the texture IH-I.

As discussed above, the most minimal Yukawa texture in this theory, which is still consistent with oscillation data, corresponds to IH-II. Due to its minimality, this scenario fails to reproduce neutrino observables within the experimental $2\sigma$ range; however, it can be fitted within $3\sigma$ values, as can be seen from the third column in Table \ref{tab:NUfit}. For this specific texture, a tension exists to simultaneously fit $\theta_{12}$ and $\theta_{23}$ close to their central values.  This attribute is demonstrated in the upper right panel  in Fig.~\ref{fig:109}. 

Moreover, for NH-I and NH-III, both   $\theta_{23}>45^\circ$ and $\theta_{23}<45^\circ$ are equally preferred (see middle left and lower panels in Fig.~\ref{fig:109}, respectively), whereas, for the texture NH-II, MCMC analysis returns solutions only for $\theta_{23}>45^\circ$ as depicted in the middle right panel  in Fig.~\ref{fig:109}.

To obtain a good fit to data, for textures with IH-I, IH-II, and NH-I, the (31)-entries in $f^L, y^L$ are required to be sizable and are of similar order compared  to other non-zero entries, as can be seen from fits Eqs.~\eqref{IH-I}--\eqref{NH-I}. Due to this requirement, the LQ masses must be much above the TeV scale to satisfy the stringent LFV processes; the most relevant process is the $\mu\to e\gamma$. The plots of this process, as a function of LQ mass, are presented in Fig.~\ref{fig:105}; they show that $M_\textrm{LQ}\gtrsim \mathcal{O}(10)$ TeV must be satisfied. On the contrary, for  textures  NH-II and NH-III, $f^L_{31}$ is set to zero, and non-zero (23)-entries are introduced. A successful fit to data requires (23)-entries being dominant, whereas  $y^L_{31}$ is somewhat small, as can be seen from fits Eqs.~\eqref{NH-II}-\eqref{NH-III}. Consequently, the branching ratio of $\mu\to e\gamma$ is highly suppressed in the latter two scenarios allowing for TeV-scale LQs. However, LQ mass below a TeV is ruled out, and the lower bound on its mass comes from the most dominating LFV processes $\mu\to eee$ and $\mu - e$ conversion, as depicted in Fig.~\ref{fig:106}. 

Further correlations among most prominent LFV processes, namely $\mu\to e\gamma$, $\mu\to eee$, and $\mu- e$ conversion are depicted in Figs.~\ref{fig:102}-\ref{fig:104}. These plots are made by marginalizing over all relevant parameters (LQ mass and Yukawa couplings) in our MCMC likelihood analysis, as discussed above. Remarkably, from these plots, it can be seen that most of the minimal textures we have exploited in this work will be entirely ruled out by upcoming low-energy experiments searching for LFV for $M_\textrm{LQ}\lesssim \mathcal{O}(100)$ TeV.  

Finally, we demonstrate how to simultaneously satisfy neutrino observables and muon $g-2$, where, currently, $(g-2)_\mu$ is the most prominent flavor anomaly that shows $4.2\sigma$ deviation from the SM prediction. As explained above, textures IH-I, IH-II, NH-I do not allow TeV scale LQs, therefore, to obtain a viable scenario, we consider an example with NH-II texture. 

NH-II consists of zero $f^L_{31}$ entry, so adding a new coupling $f^R_{32}$, for instance, will not induce a new contribution to $\mu\to e\gamma$ arising from chirality-enhanced term; it only induces a weaker $\tau \to \mu\gamma$ process. However, the benchmark provided in Eq.~\eqref{NH-II} is still unsuitable for incorporating $(g-2)_\mu$ with a TeV-scale LQ. This particular fit has a somewhat small $f^L_{32}$ element; thus, to reproduce $(g-2)_\mu$, an order unity $f^R_{32}$ coupling is needed. Once this required size of $f^R_{32}$ is included, along with the $f^L_{33}$ coupling present in Eq.~\eqref{NH-II}, top-quark chirality-enhanced contribution to $\tau\to \mu \gamma$ rate becomes too large and rules out this particular fit. Because of that, we perform a new fit by including  $(g-2)_\mu$ observable along with LFV rates in the $\chi^2$-function to allow for a TeV scale LQ mass.  From Eq.~\eqref{eq:AMM}, one can approximate  the $(g-2)_\mu$ up to the leading order as
\begin{align}
\Delta a_\mu\simeq -\frac{3}{8\pi^2} \frac{m_tm_\mu}{M^2_{LQ}} \left[ y_{32}^L y^R_{32} \left( \frac{7}{6}+\frac{2}{3}\ln \frac{m^2_t}{M^2_{LQ}} \right) - f^L_{32}f^R_{32} \left( \frac{1}{6} + \frac{2}{3} \ln \frac{m_t^2}{M^2_{LQ}} \right) \right].
\end{align}
We obtain the following parameters from the numerical fit, i.e.,
\begin{align}
&m_0=11.388\;\textrm{eV}, \; M_\textrm{LQ}=2.1\;\textrm{TeV},\; f^R_{32}=0.18,
\\
&f^L=
\left(
\begin{array}{ccc}
 0 & 0 & 0 \\
 0 & 0 & -0.870445 \\
 0 & -0.0319843 & -0.00376365 \\
\end{array}
\right),\;\;\;  
y^L=\left(
\begin{array}{ccc}
 0 & 0 & 0 \\
 0 & 0 & -0.175317 \\
 0.0248746 & -0.0344455 & -0.0702396 \\
\end{array}
\right).\label{muonFIT}
\end{align}
We verified that introduction of non-zero $f^R_{32}$ still provides sub-leading contribution to the neutrino mass, and its effect can be safely neglected. However, it has significant effect on cLFV, which we have also incorporated.  The above parameters provide a good fit to the neutrino observables
\begin{align}
&\Delta m^2_{21}= 7.414\times 10^{-5} \textrm{eV}^2,\;   \Delta m^2_{31}=2.511 \times 10^{-3} \textrm{eV}^2, 
\\
&\sin^2\theta_{12}= 0.305,\;\sin^2\theta_{23}=  0.574,\;\sin^2\theta_{13}= 0.02223,\;
\end{align}
as well as to the muon $g-2$
\begin{align}
\Delta a_\mu=   2.62\times 10^{-9}. 
\end{align}

Since top-quark chirality enhancement is required to fit  $(g-2)_\mu$ consistently with sizable $f^{L,R}_{32}$ entries,  as can be seen from Eq.~\eqref{muonFIT} that (33)-entry must be pretty small compared to the rest of the elements to keep  $\tau$ decays under control. In addition, $\mu - e$ conversion in the gold nucleus also lies just below the current bound. Branching ratios of these two leading processes for this fit are found to be
\begin{align}
&BR(\tau\to \mu\gamma)= 1.1\times 10^{-8},\;\;\;
BR(\mu - e)\; \textrm{conv.}= 6.1\times 10^{-13}.\label{muTOeValue}
\end{align}

On the other hand, the consistency with the recent lattice results that weakens the long-standing discrepancy in $(g-2)_\mu$ between experiment and theory can be obtained by reducing the value of $f^R_{32}$ without affecting the neutrino observables. For example, setting $f^R_{32}=0.134$ instead of $0.18$ leads to $\Delta a_\mu= 1.95\times 10^{-9}$ in agreement with lattice result~\cite{Borsanyi:2020mff}. Such a reduced value of this coupling subsequently decreases $\tau\to \mu\gamma$ rate; this new value of $f^R_{32}$ corresponds to $BR(\tau\to \mu\gamma)= 5.8\times 10^{-9}$, whereas $BR(\mu - e)$ conversion as quoted in Eq.~\eqref{muTOeValue} remains unaltered since this coupling plays no role for this observable.

\subsection{Non-standard Neutrino Interactions}
The LQs $R_2$ and $S_1$ couple to neutrinos and quarks (cf. Eq.~\eqref{new-yuk}), consequently, charged-current non-standard interactions (NSI) at tree-level can be induced~\cite{Wolfenstein:1977ue,Proceedings:2019qno,Babu:2019mfe}. 
Using the effective dimension-6 operators for NSI introduced in Ref.~\cite{Wolfenstein:1977ue}, the effective NSI parameters in our model can be written (in the up-quark mass diagonal basis) as, 
\begin{align}
    \varepsilon_{\alpha\beta} \ = \ \frac{3}{4\sqrt 2 G_F}
    \left(\frac{f^{L\star}_{u\alpha}f^L_{u\beta}}{M_{R^{2/3}}^2}
    +\frac{\hat{y}^L_{d\alpha}\hat{y}^L_{d\beta}}{M_{S^{1/3}}^2}
    \right) \, ,
    \label{eq:NSI}
\end{align}
where $\hat{y}^L \equiv -V^T y^L$. Note that any nonzero $y^L_{d\alpha}$ is in conjunction to Cabibbo rotation and induces $\hat{y}^L_{se}$ leading to strong constraints, for instance, $K^+ \to \pi^+ \nu \nu$ with Re[$\hat{y}^L_{de} \hat{y}^L_{se}$] $= [-3.7, 8.3] \times 10^{-4} \left(M_{S_1}/\text{TeV}\right)^2$. Thus, NSI induced from $S_1$ LQ via $y^L$ Yukawa coupling is subdominant. Moreover, any Yukawa couplings to electron and muon sector $f^L_{u\alpha}$ and $y^L_{d\alpha}$ ($\alpha= e, \mu$) are subjected to stringent constraints from the non-resonant dilepton searches~\cite{ Babu:2020hun, Angelescu:2021lln} at the LHC. 
However, the LHC limits on the LQ Yukawa coupling in the tau sector are weaker and in principle be $\mathcal{O} (1)$ leading to $\epsilon_{\tau \tau}$ as large as 34.4 \% \cite{Babu:2019mfe}, which is within reach of long-baseline neutrino experiments, such as DUNE \cite{Chatterjee:2021wac}.

\section{Conclusions}\label{sec:Con}
Neutrino oscillations were discovered almost 25 years ago, showing that neutrinos have a mass; however, its origin remains unknown. Recently, several pieces of evidence of lepton flavor universality violation strongly indicate physics beyond the SM. Scalar leptoquarks are the prime candidates for resolving all these flavor anomalies. Motivated by this,  in this work, we hypothesized that neutrino masses and flavor anomalies have a common new physics origin and proposed a new two-loop neutrino mass model consisting of scalar leptoquarks $(\overline{3},1,1/3)$ and $(3,2,7/6)$ along with a third scalar $(3,3,2/3)$. Each of these scalar leptoquarks has the potential to incorporate $R_{D^{(\ast)}}$, $(g-2)_e$, and $(g-2)_\mu$ anomalies. The scalar leptoquark $(3,2,7/6)$  may also address anomalies in the $R_{K^{(\ast)}}$ ratios via new physics interactions with the electron. Since resolution to flavor anomalies requires TeV scale scalar leptoquarks with some of the Yukawa couplings of order unity, the proposed model can be tested in ongoing and future colliders. However, probes of lepton flavor violation in neutrino mass models provide the most efficient way of searching for physics beyond the Standard Model that expands far beyond the reach of colliders such as the LHC. In this work, we have primarily focused on the neutrino phenomenology and examined various minimal textures of the Yukawa coupling matrices that can satisfy the neutrino oscillation data. In particular, we have exploited five benchmark scenarios with a limited number of Yukawa parameters of a similar order, two of which provide an inverted hierarchy for the neutrino masses, and the rest provide a normal hierarchy. Moreover, by performing a detailed numerical procedure, namely, the Markov chain Monte Carlo analysis, we have studied in depth various lepton flavor violating processes and constrained the parameter space of this theory. Our analysis shows that for the minimal Yukawa textures considered in this work, the current low-energy experiments provide stringent constraints on model parameters, and near-future experiments hunting for lepton flavor violating rare processes will rule out these scenarios for leptoquark masses below 100 TeV. Finally, we have presented a case study where neutrino observables and the tension in the muon anomalous magnetic moment, the most prominent flavor anomaly, are incorporated simultaneously for TeV scale leptoquark masses by keeping lepton flavor violations under control, which requires a bit of tuning of the Yukawa parameters.

\section*{Acknowledgments}
The work of J.J. was supported in part by the National Research and Innovation Agency of the Republic of Indonesia via Research Support Facility Program. 

\bibliographystyle{style}
\bibliography{references}
\end{document}